\documentclass[MSNbibl,nameyear,dvips]{arxstspdf}
\usepackage{graphicx}
\usepackage{flushend}
\usepackage{stfloats}

\onecolumn

\volume{26}
\issue{3}
\pubyear{2011}
\firstpage{332}
\lastpage{351}
\doi{10.1214/11-STS365}

\makeatletter
\newcommand{\binom}[2]{\pmatrix{#1\cr #2}}

\newtheorem{proposition}{Proposition}
\newtheorem{theorem}{Theorem}
\newproclaim{remark}{Remark}

\newcommand{\norm}[1]{\|#1\|}
\newcommand{\X}{{\mathsf{X}}}
\newcommand{\z}{{\mathbf{z}}}
\newcommand{\w}{{\mathbf{w}}}
\newcommand{\p}{{\mathbf{p}}}
\newcommand{\bt}{{\bolds{\theta}}}
\newcommand{\Y}{{\mathsf{Y}}}
\newcommand{\Sp}{\operatorname{Sp}}

\renewcommand{\emptyset}{\varnothing}
\newcommand{\eqref}[1]{(\ref{#1})}

\makeatother

\begin{document}
\begin{frontmatter}

\title{Improving~the~Convergence~Properties~of \mbox{the~Data~Augmentation~Algorithm}~with~an
Application~to~Bayesian~Mixture~Modeling}
\runtitle{Improving the Data Augmentation Algorithm}

\begin{aug}
\author[a]{\fnms{James P.} \snm{Hobert}\corref{}\ead[label=e1]{jhobert@stat.ufl.edu}},
\author[b]{\fnms{Vivekananda} \snm{Roy}\ead[label=e2]{vroy@iastate.edu}}
\and
\author[c]{\fnms{Christian P.} \snm{Robert}\ead[label=e3]{xian@ceremade.dauphine.fr}}
\runauthor{J. P. Hobert, V. Roy and C. P. Robert}

\affiliation{University of Florida, Iowa State University and Universit\'{e} Paris Dauphine \& CREST, INSEE}

\address[a]{James P. Hobert is Professor, Department of Statistics,
University of
Florida,
221 Griffin--Floyd Hall, P.O. Box 118545,
 Gainesville, Florida 32611, USA \printead{e1}.}
\address[b]{Vivekananda Roy is Assistant Professor, Department of Statistics, Iowa
State University,
3415 Snedecor Hall,
Ames, Iowa 50011, USA \printead{e2}.}
\address[c]{Christian P. Robert is Professor, CEREMADE,
Universit\'{e} Paris-Dauphine, 75775 Paris cedex 16, Senior Member of Institut Universitaire de
France and Senior Researcher, CREST, 92245 Malakoff cedex, France \printead{e3}.}

\end{aug}

%
\begin{abstract}
The reversible Markov chains that drive the data augmentation (DA)
and sandwich algorithms define self-adjoint operators whose spectra
encode the convergence properties of the algorithms. When the
target distribution has uncountable support, as is nearly always the
case in practice, it is generally quite difficult to get a handle on
these spectra. We show that, if the augmentation space is finite,
then (under regularity conditions) the operators defined by the DA
and sandwich chains are compact, and the spectra are finite subsets
of $[0,1)$. Moreover, we prove that the spectrum of the sandwich
operator dominates the spectrum of the DA operator in the sense that
the ordered elements of the former are all less than or equal to the
corresponding elements of the latter. As a concrete example, we
study a widely used DA algorithm for the exploration of posterior
densities associated with Bayesian mixture models
[\textit{J. Roy. Statist. Soc. Ser. B}
\textbf{56} (1994) 363--375]. In particular, we compare this mixture DA
algorithm with an alternative algorithm proposed by
Fr\"uhwirth-Schnatter [\textit{J. Amer. Statist. Assoc.} \textbf{96} (2001)
194--209] that is based on random label switching.
\end{abstract}

%

\end{frontmatter}

\section{Introduction}
\label{secintro}

Suppose that $f_X \dvtx  \mathbb{R}^p \rightarrow[0,\infty)$ is a
probability density function that is intractable in the sense that
expectations with respect to $f_X$ cannot be computed analytically.
If direct simulation from $f_X$ is infeasible, then classical Monte
Carlo methods cannot be used to explore $f_X$ and one might resort to
a~Markov chain Monte Carlo (MCMC) method such as the data augmentation
(DA) algorithm (Tanner and Wong, \citeyear{tannwong1987};
Liu, Wong and Kong, \citeyear{liuwongkong1994};
Hobert, \citeyear{hobe2011}).
To build a DA algorithm, one must identify a joint density, say,
$f\dvtx
\mathbb{R}^p \times\mathbb{R}^q \rightarrow[0,\infty)$, that
satisfies two conditions: (i) the $x$-marginal of $f(x,y)$ is $f_X$,
and (ii) sampling from the associated conditional densities,
$f_{X|Y}(\cdot|y)$ and $f_{Y|X}(\cdot|x)$, is straightforward. (The
$y$-coordinate may be discrete or continuous.) The first of the two
conditions allows us to construct a Markov chain having $f_X$ as an
invariant density, and the second ensures that we are able to simulate
this chain. Indeed, let $\{X_n\}_{n=0}^\infty$ be a Markov chain
whose dynamics are defined (implicitly) through the following two-step
procedure for moving from the current state, $X_n=x$, to $X_{n+1}$ (see Procedure~\ref{proc1}).

It is well known and easy to establish that the DA Markov
chain is reversible with respect to $f_X$, and this of course implies
that $f_X$ is an invariant density (Liu, Wong and Kong, \citeyear
{liuwongkong1994}).
Consequently, if the chain satisfies the usual regularity conditions
(see Section~\ref{secot}), then we can use averages to consistently
estimate intractable expectations with respect to $f_X$
(Tierney, \citeyear{tier1994}). The resulting MCMC algorithm is known
as a DA
algorithm for $f_X$. (Throughout this section, $f_X$ is assumed to be
a probability density function, but, starting in Section~\ref{secot}, a
more general version of the problem is considered.)

\def\tablename{Procedure}
\begin{table}
\caption{Iteration $n+1$ of the DA Algorithm}\label{proc1}
\begin{tabular}{@{}p{240pt}@{}}
\hline\vspace*{-13pt}
\begin{enumerate}[2.]
\item Draw $Y \sim f_{Y|X}(\cdot|x)$, and call the observed value $y$.
\item Draw $X_{n+1} \sim f_{X|Y}(\cdot|y)$.
\end{enumerate}
\mbox{}\vspace*{-18pt}\\
\hline
\end{tabular}
\end{table}

When designing a DA algorithm, one is free to choose any joint density
that satisfies conditions (i) and~(ii). Obviously, different joint
densities will yield different DA chains, and the goal is to find a
joint density whose DA chain has good convergence properties. (This
is formalized in Section~\ref{secDA} using $\chi^2$-distance to
stationarity.) Unfortunately, the ``ideal'' joint density, which
yields the DA chain with the fastest possible rate of convergence,
does not satisfy the simulation requirement. Indeed, consider
$f_{\perp}(x, y) = f_X(x)   g_Y(y)$, where $g_Y(y)$ is any density
function on $\mathbb{R}^q$. Since $f_{\perp}(x,y)$ factors,
$f_{X|Y}(x|y) = f_X(x)$ and it follows that the DA chain is just an
i.i.d. sequence from $f_X$. Of course, this ideal DA algorithm is
useless from a practical standpoint because, in order to simulate the
chain, we must draw from $f_X$, which is impossible. We return to
this example later in this section.

It is important to keep in mind that there is no inherent interest in
the joint density $f(x,y)$. It is merely a tool that facilitates
exploration of the target density, $f_X(x)$. This is the reason why
the DA chain does not possess a $y$-coordinate. In contrast, the
two-variable Gibbs sampler based on $f_{X|Y}(\cdot|y)$ and
$f_{Y|X}(\cdot|x)$, which is used to explore $f(x,y)$, has both $x$
and $y$-coordinates. So, while the two-step procedure described above
can be used to simulate both the DA and Gibbs chains, there is one key
difference. When simulating the DA chain, we do not keep track of the
$y$-coordinate.

Every reversible Markov chain defines a self-adjoint operator whose
spectrum encodes the convergence properties of the chain
(Mira and Geyer, \citeyear{mirageye1999};
Rosenthal, \citeyear{rose2003};
Diaconis, Khare and Saloff-Coste, \citeyear{diackharsalo2008}). Let $X
\sim
f_X$ and consider the space of functions $g$ such that the random
variable $g(X)$ has finite variance and mean zero. To be more
precise, define
\begin{eqnarray*}
L^2_0(f_X) =  \biggl\{ g\dvtx  \mathbb{R}^p \rightarrow\mathbb{R} \dvtx
\int_{\mathbb{R}^p} g^2(x)   f_X(x) \, dx < \infty \mbox{ and }
\int_{\mathbb{R}^p} g(x)   f_X(x) \, dx = 0  \biggr\}  .
\end{eqnarray*}
Let $k(x'|x)$ be the Markov transition density (Mtd) of the DA chain.
(See Section~\ref{secDA} for a formal definition.) This Mtd defines
an operator, $K\dvtx  L^2_0 \rightarrow L^2_0 $, that maps $g(x)$ to
\[
(Kg)(x) := \int_{\mathbb{R}^p} g(x')   k(x'|x) \, dx'  .
\]
Of course, $(Kg)(x)$ is just the expected value of $g(X_1)$ given that
$X_0=x$. Let $I\dvtx  L^2_0 \rightarrow L^2_0$ denote the identity
operator, which leaves functions unaltered, and consider the operator
$K-\lambda I$, where $\lambda\in\mathbb{R}$. By definition,
$K-\lambda I$ is \textit{invertible} if, for each $h \in L^2_0$, there
exists a unique $g \in L^2_0$ such that $((K-\lambda I)g)(x) = (Kg)(x)
- \lambda g(x) = h(x)$. The spectrum of $K$, which we denote by
$\Sp(K)$, is simply the set of $\lambda$ such that $K-\lambda I$ is
\textit{not} invertible. Becau\-se~$K$ is defined through a DA chain,
$\Sp(K) \subseteq[0,1]$ (see Section~\ref{secDA}). The number of
elements in $\Sp(K)$ may be finite, countably infinite or uncountable.

In order to understand what ``good'' spectra look like, consider the
ideal DA algorithm introduced earlier. Let $k_{\perp}$ and
$K_{\perp}$ denote the Mtd and the corresponding operator,
respectively. In the ideal case, $X_{n+1}$ is independent of $X_n$
and has density $f_X$. Therefore, the Mtd is just
$k_{\perp}(x'|x)=f_X(x')$ and
\[
(K_{\perp}g)(x) = \int_{\mathbb{R}^p} g(x')   f_X(x') \, dx' = 0  ,
\]
which implies that
\[
\bigl((K_{\perp}-\lambda I)g\bigr)(x) = -\lambda g(x)  .
\]
It follows that $K_{\perp}\,{-}\,\lambda I$ is invertible as long as
\mbox{$\lambda\,{\ne}\,0$}. Hence, the ``ideal spectrum'' is $\Sp(K_{\perp}) =
\{0\}$. Loosely speaking, the closer $\Sp(K)$ is to $\{0\}$, the
faster the DA algorithm converges (Diaconis, Khare and Saloff-Coste,
\citeyear{diackharsalo2008}).

Unfortunately, in general, there is no simple me\-thod for calculating
$\Sp(K)$. Even getting a handle on $\Sp(K)$ is currently difficult.
However, there is one situation where $\Sp(K)$ has a very simple
structure. Let $\Y=  \{y \in\mathbb{R}^q \dvtx  f_Y(y) > 0  \}$,
where $f_Y(y) = \int_{\mathbb{R}^p} f(x,y) \, dx$. We show that when
$\Y$ is a finite set, $\Sp(K)$ consists of a finite number of elements
that are directly related to the Markov transition matrix (Mtm) of the
so-called conjugate chain, which is the reversible Markov chain that
lives on $\Y$ and makes the transition $y \rightarrow y'$ with
probability $\int_{\mathbb{R}^p} f_{Y|X}(y'| x)   f_{X|Y}(x|y) \, dx$.
In particular, we prove that when $|\Y|=d<\infty$, $\Sp(K)$ consists
of the point $\{0\}$ together with the $d-1$ smallest eigenvalues of
the Mtm of the conjugate chain. We use this result to prove that the
spectrum associated with a particular alternative to the DA chain is
closer than $\Sp(K)$ to the ideal spectrum, $\{0\}$.

\begin{table}
\caption{Iteration $n+1$ of the Sandwich Algorithm}\label{proc2}
\begin{tabular}{@{}p{240pt}@{}}
\hline\vspace*{-13pt}
\begin{enumerate}[3.]
\item Draw $Y \sim f_{Y|X}(\cdot|x)$, and call the observed value $y$.
\item Draw $Y' \sim r(\cdot|y)$, and call the observed value $y'$.
\item Draw $\tilde{X}_{n+1} \sim f_{X|Y}(\cdot|y')$.
\end{enumerate}
\mbox{}\vspace*{-18pt}\\
\hline
\end{tabular}
\end{table}

DA algorithms often suffer from slow convergence, which is not
surprising given the close connection between DA and the notoriously
slow to converge EM algorithm (see, e.g., van Dyk and Meng, \citeyear
{vandmeng2001}).
Over the last decade, a great deal of effort has gone into modifying
the DA algorithm to speed convergence. See, for example,
Meng and van Dyk (\citeyear{mengvand1999}), Liu and Wu (\citeyear
{liuwu1999}), Liu and Sabatti (\citeyear{liusaba2000}),
van Dyk and Meng (\citeyear{vandmeng2001}), Papaspiliopoulos, Roberts
and Sk\"old (\citeyear{paparobeskol2007}),
Hobert and Marchev (\citeyear{hobemarc2008}) and Yu and Meng
(\citeyear{yumeng2011}). In this paper we
focus on the so-called sandwich algorithm, which is a simple
alternative to the DA algorithm that often converges much faster. Let
$r(y'|y)$ be an auxiliary Mtd (or Mtm) that is reversible with respect
to $f_Y$, and consider a new Markov chain,
$\{\tilde{X}_n\}_{n=0}^\infty$, that moves from $\tilde{X}_n=x$ to
$\tilde{X}_{n+1}$ via the following \textit{three-step} procedure (see Procedure~\ref{proc2}).

A routine calculation shows that the sandwich chain remains
reversible with respect to $f_X$, so it is a~viable alternative to the
DA chain. The name ``sandwich algorithm'' was coined by
Yu and Meng (\citeyear{yumeng2011}) and is based on the fact that the
extra draw from
$r(\cdot|y)$ is sandwiched between the two steps of the DA algorithm.
Clearly, on a per iteration basis, it is more expensive to simulate
the sandwich chain. However, it is often possible to find an $r$ that
leads to a substantial improvement in mixing despite the fact that it
only provides a low-dimensional (and hence inexpensive) perturbation
on the $\Y$ space. In fact, the computational cost of drawing from
$r$ is often negligible relative to the cost of drawing from
$f_{Y|X}(\cdot|x)$ and~$f_{X|Y}(\cdot|y)$. Concrete examples can be
found in Meng and van Dyk (\citeyear{mengvand1999}), Liu and Wu
(\citeyear{liuwu1999}),
van Dyk and Meng (\citeyear{vandmeng2001}), Roy and Hobert (\citeyear
{royhobe2007}) and
Section~\ref{secapp} of this paper.

Let $\tilde{k}(x'|x)$ denote the Mtd of the sandwich chain. Also, let
$\tilde{K}$ and $\Sp(\tilde{K})$ denote the corresponding operator and
its spectrum. The main theoretical result in this paper provides
conditions under which~$\Sp(\tilde{K})$ is closer than~$\Sp(K)$ to the
ideal spectrum. Recall that when $|\Y|=d<\infty$, $\Sp(K)$ consists
of the point~$\{0\}$ and the $d-1$ smallest eigenvalues of the Mtm of
the conjugate chain. If, in addition, $r$ is idempotent (see
Section~\ref{secimprove} for the definition), then $\Sp(\tilde{K})$
consists of the point $\{0\}$ and the $d-1$ smallest eigenvalues of
a~\textit{different} $d \times d$ Mtm, and $0 \le\tilde{\lambda}_i
\le
\lambda_i$ for all $i \in\{1,2,\ldots,d-1\}$, where
$\tilde{\lambda}_i$ and $\lambda_i$ are the $i$th largest elements of
$\Sp(\tilde{K})$ and $\Sp(K)$, respectively. So $\Sp(\tilde{K})$
dominates $\Sp(K)$ in the sense that the ordered elements of
$\Sp(\tilde{K})$ are uniformly less than or equal to the corresponding
elements of $\Sp(K)$. We conclude that the sandwich algorithm is
closer than the DA algorithm to the gold standard of classical Monte
Carlo.

One might hope for a stronger result that quantifies the extent to
which the sandwich chain is better than the DA chain, but such a
result is impossible without further assumptions. Indeed, if we take
the auxiliary Markov chain on $\Y$ to be the degenerate chain that is
absorbed at its starting point, then the sandwich chain is the same as
the DA chain.

To illustrate the huge gains that are possible through the sandwich
algorithm, we introduce a new example involving a Bayesian mixture
model. Let $Z_1,\ldots,Z_m$ be a random sample from a $k$-component
mixture density taking the form
%
\begin{equation}
\label{eqmix}
\sum_{j=1}^k p_j h_{\theta_j}(z)  ,
\end{equation}
where $\theta_1,\ldots,\theta_k \in\Theta\subseteq\mathbb{R}^l$,
$\{h_\theta(\cdot) \dvtx  \theta\in\Theta \}$ is a parametric family
of densities, and the $p_j$'s are non-negative weights that sum to one.
Of course, a Baye\-sian analysis requires priors for the unknown
parameters, which are $\bt= (\theta_1, \theta_2,\ldots, \theta_k)^T$
and $\p= (p_1, p_2,\ldots,  p_k)^T$. In typical applications we have
no prior information on~$p$, and the same (lack of) prior information
about each of the components in the mixture. Thus, it makes sense to
put a symmetric Dirichlet prior on the weights, and to take a prior
on~$\bt$ that has the form $\prod_{j=1}^k \pi(\theta_j)$, where $\pi\dvtx
\Theta\rightarrow[0,\infty)$ is a proper prior density on $\Theta$.
Let $\z= (z_1,\ldots, z_m)$ denote the observed data. It is well known
that the resulting posterior density, $\pi(\bt,\p|\z)$, is intract\-able
and highly multi-modal (see, e.g., Jasra, Holmes and Stephens,
\citeyear{jasrholmstep2005}). Indeed, let $E$ denote any one of
the $k!$ permutation matrices of dimension $k$ and note that
$\pi(\bt,\p|\z) = \pi(E \bt, E \p|\z)$. Thus, every local
maximum of
the posterior density has $k!-1$ exact replicas somewhere else in the
parameter space.

The standard DA algorithm for this mixture problem was introduced by
Diebolt and Robert (\citeyear{diebrobe1994}) and is based on the
following augmented model.
Assume that $\{(Y_i,Z_i)\}_{i=1}^m$ are i.i.d. pairs such that $Y_i=j$
with probability $p_j$, and, conditional on $Y_i=j$, $Z_i \sim
h_{\theta_j}(\cdot)$. Note that the marginal density of $Z_i$ under
this two-level hierarchy is just \eqref{eqmix}. Let ${\mathbf{y}}=
(y_1,\ldots,y_m)$ denote a realization of the $Y_i$'s. The so-called
complete data posterior density, $\pi((\bt,\p),{\mathbf{y}}|\z)$,
is just the
posterior density that results when we combine our model for
$\{(Y_i,Z_i)\}_{i=1}^m$ with the priors on $\p$ and $\bt$ defined
above. It is easy to see that
\[
\sum_{{\mathbf{y}}\in\Y} \pi( (\bt,\p),{\mathbf{y}}|\z) = \pi
(\bt,\p|\z)  ,
\]
where $\Y$ is the set of all sequences of length $m$ consisting of
integers from the set $\{1,\ldots,k\}$. Hence, $\pi((\bt,\p
),{\mathbf{y}}|\z)$
can be used to build a DA algorithm as long as it is possible to
sample from the conditionals, $\pi((\bt,\p)|{\mathbf{y}},\z)$ and
$\pi({\mathbf{y}}|(\bt,\p),\z)$. We call it the mixture DA (MDA) algorithm.
Note that the state space for the MDA chain is the Cartesian product
of~$\mathbb{R}^{kl}$ and the $k$-dimensional simplex, but
$|\Y| = k^m < \infty$.

The MDA algorithm often converges very slowly because it moves between
the symmetric modes of $\pi(\bt,\p|\z)$ too infrequently
(Celeux, Hurn and Robert, \citeyear{celehurnrobe2000};
Lee et al., \citeyear{leemarimengrobe2008}).
Fr\"uhwirth-Schnatter (\citeyear{fruh2001}) suggested adding a random
label switching step to
each iteration of the MDA algorithm in order to force movement between
the modes. We show that the resulting Markov chain, which we call
the FS chain, is a special case of the sandwich chain. Moreover, our
theoretical results are applicable and imply that the spectrum of the
operator defined by the FS chain dominates the spectrum of the MDA
operator. To illustrate the extent to which the label switching step
can speed convergence, we study two specific mixture models and
compare the spectra associated with the FS and MDA chains. The first
example is a~toy problem in which we are able to get exact formulas
for the eigenvalues. The second example is a~normal mixture model
that is frequently used in practice, and we approximate the
eigenvalues via classical Monte Carlo methods. The conclusions from
the two examples are quite similar. First, the MDA chain converges
slowly and its rate of convergence deteriorates very rapidly as the
sample size, $m$, increases. Second, the FS chain converges much
faster and its rate does not seem as adversely affected by increasing
sample size.

The remainder of this paper is organized as follows.
Section~\ref{secot} is a brief review of the operator theory used for
analyzing reversible Markov chains. Section~\ref{secDA} contains a
string of results about the DA operator and its spectrum. Our main
result comparing the DA and sandwich chains in the case where
$|\Y|<\infty$ appears in Section~\ref{secimprove}.
Section~\ref{secapp} contains a~detailed review of the MDA and FS
algorithms, as well as a~proof that the FS chain is a~special case of
the sandwich chain. Finally, in Section~\ref{secexamples}, the MDA
and FS chains are compared in the context of two specific examples.
The \hyperref[appm]{Appendix} contains an eigen-analysis of a special $4 \times4$ Mtm.

\section{Operator Theory for Reversible Markov Chains}
\label{secot}

Consider the following generalized version of the problem described in
the \hyperref[secintro]{Introduction}. Let $\X$ be a~general space (equipped with a
countably generated $\sigma$-algebra) and suppose that $f_X\dvtx  \X
\rightarrow[0,\infty)$ is an intractable probability density with
respect to the measure $\mu$. Let $p(x'|x)$ be a Mtd (with respect to
$\mu$) such that $p(x'|x) f_X(x)$ is symmetric in $(x,x')$, so the
Markov chain defined by $p$ is reversible with respect to~$f_X(x)$.
Assume that the chain is Harris ergodic, which means that it is
irreducible, aperiodic and Harris recurrent
(Meyn and Tweedie, \citeyear{meyntwee1993};
Asmussen and Glynn, \citeyear{asmuglyn2010}).

Define the Hilbert space
\begin{eqnarray*}
L^2_0(f_X) =  \biggl\{ g\dvtx  \X\rightarrow\mathbb{R} \dvtx  \int_\X g^2(x)
f_X(x) \mu(dx) < \infty
\mbox{ and }   \int_\X g(x) f_X(x) \mu(dx)
= 0  \biggr\}  ,
\end{eqnarray*}
where inner product is defined as
\[
\langle g,h \rangle= \int_\X g(x)   h(x)   f_X(x)
\mu(dx)  .
\]
The corresponding norm is given by $\norm{g} = \sqrt{\langle g,g
\rangle}$. The Mtd $p$ defines an operator $P\dvtx  L^2_0(f_X)
\rightarrow L^2_0(f_X)$ that acts on $g \in L^2_0(f_X)$ as follows:
\[
(Pg)(x) = \int_\X g(x')   p(x'|x)   \mu(dx')  .
\]
It is easy to show, using reversibility, that for $g, h \in
L^2_0(f_X)$, $\langle Pg,h \rangle= \langle g,Ph \rangle$; that is,
$P$ is a self-adjoint operator. The spectrum of $P$ is defined as
\[
\Sp(P) =  \{ \lambda\in\mathbb{R}\dvtx  P - \lambda I   \mbox{ is
not invertible}  \}  .
\]
There are two ways in which $P - \lambda I$ can fail to be invertible
(Rudin, \citeyear{rudi1991}, Chapter 4). First, $P - \lambda I$ may
not be
onto, that is, if there exists $h \in L^2_0(f_X)$ such that there is
no $g \in L^2_0(f_X)$ for which \mbox{$((P\,{-}\,\lambda I)g)\,{=}\,h$}, then the range
of $P-\lambda I$ is not all of $L^2_0(f_X)$, so $P-\lambda I$ is not
invertible and $\lambda\in\Sp(P)$. Second, $P-\lambda I$ may not
be one-to-one, that is, if there exist two different functions $g, h
\in L^2_0(f_X)$ such that $((P-\lambda I)g) = ((P-\lambda I)h)$, then
$P-\lambda I$ is not one-to-one, so $P-\lambda I$ is not invertible
and $\lambda\in\Sp(P)$. Note that if $((P-\lambda I)g) =
((P-\lambda I)h)$, then $Pg^* = \lambda g^*$ with $g^*=g-h$, and
$\lambda$ is called an eigenvalue with eigen-function $g^*$. We call
the pair $(\lambda,g^*)$ an eigen-solution.

Let $L^2_{0,1}(f_X)$ denote the subset of\vspace*{1pt} functions in $L^2_0(f_X)$
that satisfy $\int_{\X} g^2(x)   f_X(x)   \mu(dx) = 1$. The
(operator) norm of $P$ is defined as
\[
\norm{P} = \sup_{g \in L^2_{0,1}(f_X)} \norm{Pg}  .
\]
A simple application of Jensen's inequality shows that the non-negative
quantity $\norm{P}$ is bounded above by 1. The norm of $P$ is a good
univariate summary of $\Sp(P)$. Indeed, define
\[
l_P = \inf_{g \in L^2_{0,1}(f_X)} \langle Pg,g \rangle
 \quad \mbox{and} \quad    u_P = \sup_{g \in L^2_{0,1}(f_X)} \langle
Pg,g \rangle .
\]
It follows from standard linear operator theory that
$\inf\Sp(P)=l_P, \sup\Sp(P)=u_P$, and $\norm{P}=\max \{-l_P,\allowbreak u_P\}$. Consequently,
 \[
 \Sp(P) \subseteq [ -\norm{P}, \norm{P}
 ] \subseteq[-1,1].
 \]
  Another name for $\norm{P}$ in this context
is the \textit{spectral radius}, which makes sense since $\norm{P}$
represents the maximum distance that $\Sp(P)$ extends away from the
origin. The quantity $1-\norm{P}$ is called the
\textit{spectral gap}.

It is well known that $\norm{P}$ is closely related to the convergence
properties of the Markov chain defined by~$p$
(Liu, Wong and Kong, \citeyear{liuwongkong1995};
Rosenthal, \citeyear{rose2003}). In particular, the chain is
geometrically ergodic if and only if $\norm{P}<1$
(Roberts and Rosenthal, \citeyear{roberose1997}). There is an
important practical advantage to
using an MCMC algorithm that is driven by a geometrically ergodic
Markov chain. Indeed, when the chain is geometric, sample averages
satisfy central limit theorems, and these allow for the computation of
asymptotically valid standard errors for MCMC-based estimates
(Jones et al., \citeyear{joneharacaffneat2006};
Flegal, Haran and Jones, \citeyear{flegharajone2008}). We note that
geome\-tric ergodicity of reversible Monte Carlo Markov chains is
typically not proven by showing that the operator norm is strictly
less than 1, but rather by establishing a so-called geometric drift
condition (Jones and Hobert, \citeyear{jonehobe2001}).

If $|\X|\,{<}\,\infty$, then $P$ is simply the Mtm whose $(i,j)$th element
is $p(j|i)$, the probability that the chain mo\-ves from $i$ to $j$. In
this case, $\Sp(P)$ is just the set of eigenvalues of $P$ (see, e.g.,
Mira and Geyer, \citeyear{mirageye1999}). The reader is probably used
to thinking of 1
as an eigenvalue for $P$ because $P$ satisfies the equation $P \mathbf{1}
= \mathbf{1}$, where $\mathbf{1}$ denotes a vector of ones. However, the
only constant function in $L_0^2$ is the zero function, so $(1,\mathbf
{ 1})$ is not a viable eigen-solution in our context. Furthermore,
irreducibility implies that the only vectors $\mathbf{v}$ that solve the
equation $P\mathbf{v}=\mathbf{v}$ are constant. It follows that $1
\notin
\Sp(P)$. Aperiodicity implies that $-1 \notin\Sp(P)$. Hence, when
$\X$ is a finite set, $\norm{P}$ is necessarily less than one. In the
next section we return to the DA algorithm.

\section{The Spectrum of the DA Chain}
\label{secDA}

Suppose that $\Y$ is a second general space and that $\nu$ is a
measure on $\Y$. Let $f\dvtx  \X\times\Y\rightarrow[0,\infty)$ be
a~joint probability density with respect to \mbox{$\mu\times\nu$}. Assume
that $\int_{\Y} f(x,y)   \nu(dy) = f_X(x)$ and that simulating from
the associated conditional densities, $f_{X|Y}(\cdot|y)$ and
$f_{Y|X}(\cdot|x)$, is straightforward. (For convenience, we assume
that $f_X$ and $f_Y$ are strictly positive on $\X$ and $\Y$,
respectively.) The DA chain, $\{X_n\}_{n=0}^\infty$, has Mtd (with
respect to $\mu$) given by
%
\begin{equation}
\label{eqMtd}
k(x'|x) = \int_{\Y} f_{X|Y}(x'|y)   f_{Y|X}(y|x)   \nu(dy)  .
\end{equation}
It is easy to see that $k(x'|x)   f_X(x)$ is symmetric in $(x,x')$,
so the DA chain is reversible with respect to~$f_X$. We assume
throughout this section and the next that all DA chains (and their
conjugates) are Harris ergodic. [See Hobert (\citeyear{hobe2011}) for
a simple
sufficient condition for Harris ergodicity of the DA chain.] If the
integral in \eqref{eqMtd} is intractable, as is nearly always the
case in practice, then direct simulation from $k(\cdot|x)$ will be
problematic. This is why the indirect two-step procedure is used.

Liu, Wong and Kong (\citeyear{liuwongkong1994}) showed that the DA
chain satisfies an
important property that results in a~positive spectrum. Let $K$
denote the operator defined by the DA chain. For $g \in L^2_0(f_X)$,
we have
\begin{eqnarray*}
\langle Kg,g \rangle &=& \int_\X(Kg)(x)   g(x)   f_X(x)
\mu(dx) \\
&=& \int_\X \biggl[ \int_\X g(x')   k(x'|x)   \mu(dx')
 \biggr] g(x)   f_X(x)   \mu(dx) \\
&=& \int_\X\biggl [ \int_\X g(x')
   \biggl[ \int_\Y f_{X|Y}(x'|y)   f_{Y|X}(y|x)   \nu(dy)  \biggr]\cdot\mu(dx')  \biggr]  g(x)   f_X(x)   \mu(dx) \\
&=& \int_\Y \biggl[
\int_\X g(x)   f_{X|Y}(x|y)   \mu(dx)  \biggr]^2    f_Y(y)
\nu(dy) \\
&\ge&0  ,
\end{eqnarray*}
which shows that $K$ is a \textit{positive operator}. It follows that
$l_K \ge0$, so $\Sp(K) \subseteq[0,\norm{K}] \subseteq[0,1]$ and
$\norm{K} = \sup\Sp(K)$.

In most applications of the DA algorithm, $f_X$ is a~probability
density function (with respect to Lebes\-gue measure), which means that
$\X$ is not finite. Typically, when $|\X|=\infty$, it is difficult to
get a handle on $\Sp(K)$, which can be quite complex and may contain
an uncountable number of points. However, if~$K$ is a~compact
operator,\footnote{The operator $K$ is defined to be compact if for any
sequence\vspace*{-1pt} of functions $g_i$ in $L_0^2(f_X)$ with $\norm{g_i} \le1$,
there is a\vspace*{-1pt}~subsequen\-ce~$g_{i_j}$ such that the sequence $Kg_{i_j}$
converges to a~limit in $L_0^2(f_X)$.} then $\Sp(K)$ has a
particularly simple form. Indeed, if $|\X|=\infty$ and $K$ is
compact, then the following all hold: (i) the number of points in
$\Sp(K)$ is at most countably infinite, (ii)~$\{0\} \in\Sp(K)$,
(iii)~$\{0\}$ is the only possible accumulation point, and (iv) any point in
$\Sp(K)$ other than~$\{0\}$ is an eigenvalue. In the remainder of
this section we prove that, if $|\X|=\infty$ and $|\Y|=d<\infty$,
then~$K$ is a compact operator and $\Sp(K)$ consists of the point~$\{0\}$
along with $d-1$ eigenvalues, and these are exactly the $d-1$
eigenvalues of the Mtm that defines the conjugate chain. It follows
immediately that the DA chain is geometrically (in fact, uniformly)
ergodic. Moreover, $K$ has a finite spectral decomposition that
provides very precise information about the convergence of the DA
chain (Diaconis, Khare and Saloff-Coste, \citeyear{diackharsalo2008}).
Indeed, let
$\{(\lambda_i,g_i)\}_{i=1}^{d-1}$ denote a set of (orthonormal)
eigen-solutions for $K$. If the chain is started at $X_0=x$, then the
$\chi^2$-distance between the distribution of $X_n$ and the stationary
distribution can be expressed as
%
\begin{equation}
\label{eqcs}
 \quad \int_\X\frac{ | k^n(x'|x) - f_X(x')  |^2}{f_X(x')}
\mu(dx') = \sum_{i=1}^{d-1} \lambda_i^{2n} g^2_i(x)  ,\hspace*{-4pt}
\end{equation}
where $k^n(\cdot|x)$ is the $n$-step Mtd, that is, the density of
$X_n$ given $X_0=x$. Of course, the $\chi^2$-distance is an upper
bound on the total variation distance (see, e.g., Liu, Wong and
Kong, \citeyear{liuwongkong1995}).
Since the $\lambda_i$'s are the
eigenvalues of the Mtm of the conjugate chain, there is some hope of
calculating, or at least bounding them.

Let $L^2_0(f_Y)$ be the set of mean-zero, square integrable functions
with respect to $f_Y$. In a slight abuse of notation, we will let
$\langle\cdot,\cdot\rangle$ and $\norm{\cdot}$ do double duty as
inner product and norm on both $L^2_0(f_X)$ and on~$L^2_0(f_Y)$. We
now describe a representation of the operator $K$ that was developed
and exploited by Diaconis, Khare and Saloff-Coste (\citeyear{diackharsalo2008})
(see also Buja, \citeyear{buja1990}). Define $Q\dvtx  L^2_0(f_X)
\rightarrow L^2_0(f_Y)$ and
$Q^*\dvtx  L^2_0(f_Y) \rightarrow L^2_0(f_X)$ as follows:
\[
(Qg)(y) = \int_\X g(x)   f_{X|Y}(x|y)   \mu(dx)
\]
and
\[
(Q^*h)(x) = \int_\Y h(y)   f_{Y|X}(y|x)
\nu(dy)  .
\]
Note that
\begin{eqnarray*}
\langle Qg,h \rangle  &=& \int_\Y(Qg)(y)   h(y)   f_Y(y) \nu(dy) \\
& =& \int_\Y \biggl[ \int_\X g(x)   f_{X|Y}(x|y)
\mu(dx)  \biggr] h(y)   f_Y(y)   \nu(dy) \\
& = &\int_\X g(x) \biggl [\int_\Y h(y)   f_{Y|X}(y|x)   \nu(dy)  \biggr] f_X(x) \mu(dx) \\
& =& \langle g,Q^*h \rangle,
\end{eqnarray*}
which shows that $Q^*$ is the adjoint of $Q$. [Note that we are using
the term ``adjoint'' in a somewhat nonstandard way since $\langle
Qg,h \rangle$ is an inner product on $L^2_0(f_Y)$, while $\langle
g,Q^*h \rangle$ is an inner product on~$L^2_0(f_X)$.] Moreover,
\begin{eqnarray*}
(Kg)(x)& =& \int_\X g(x')   k(x'|x)   \mu(dx') \\
& =& \int_\X
g(x')  \biggl[ \int_\Y f_{X|Y}(x'|y)   f_{Y|X}(y|x)   \nu(dy)  \biggr]
\mu(dx') \\
& = &\int_\Y \biggl[ \int_\X g(x')   f_{X|Y}(x'|y)
\mu(dx')  \biggr] f_{Y|X}(y|x)   \nu(dy) \\
& =& \int_\Y(Qg)(y)
f_{Y|X}(y|x)   \nu(dy) \\
&  = &((Q^*Q)g)(x)  ,
\end{eqnarray*}
which shows that $K = Q^*Q$. As in Section~\ref{secintro}, consider
the conjugate Markov chain whose Mtd (with respect to $\nu$) is given
by
%
\begin{equation}
\label{eqMtdy}
\hat{k}(y'|y) = \int_{\X} f_{Y|X}(y'|x)   f_{X|Y}(x|y)   \mu(dx)
 .
\end{equation}
Obviously, $\hat{k}(y'|y)$ is reversible with respect\vspace*{1pt}
to $f_Y$. Fur\-thermore, it is easy to see that $\hat{K} = QQ^*$, where $\hat{K}\dvtx
L^2_0(f_Y) \rightarrow L^2_0(f_Y)$ is the operator associated with~$\hat{k}$.

Now suppose that $(\lambda,g)$ is an eigen-solution for~$K$, that is,
$(Kg)(x) = \lambda g(x)$, which is equivalent to $((Q^*Q)g)(x) = \lambda
g(x)$. Applying the operator $Q$ to both sides yields
$(Q((Q^*Q)g))(y) = \lambda(Qg)(y)$, but we can rewrite this as
$(\hat{K}(Qg))(y) = \lambda(Qg)(y)$, which shows that $(\lambda, Qg)$
is an eigen-solution for $\hat{K}$. [See Buja (\citeyear{buja1990})
for a
similar development.] Of course, the same argument can be used to
convert an eigen-solution for $\hat{K}$ into an eigen-solution for
$K$. We conclude that $\hat{K}$ and $K$ share the same eigenvalues.
Here is a~precise statement.

\begin{proposition}
\label{propsameevs}
If $(\lambda,g)$ is an eigen-solution for~$K$, then $(\lambda,(Qg))$
is an eigen-solution for $\hat{K}$. Conversely, if $(\lambda,h)$ is
an eigen-solution for $\hat{K}$, then $(\lambda,(Q^*h))$ is an
eigen-solution for $K$.
\end{proposition}

\begin{remark}
Diaconis, Khare and Saloff-Coste (\citeyear{diackharsalo2008})
describe several examples where the
eigen-solutions of $K$ and $\hat{K}$ can be calculated explicitly.~The\-se authors
studied the case where $f_{X|Y}(x|y)$ is an~uni\-variate
exponential family (with $y$ playing the~role of the parameter), and
$f_Y(y)$ is the conjugate~prior.
\end{remark}

The next result, which is easily established using minor extensions of
results in Retherford's (\citeyear{reth1993}) Chapter VII, shows that compactness is a
solidarity property for $K$ and $\hat{K}$.

\begin{proposition}
\label{propsamecompact}
$K$ is compact if and only if $\hat{K}$ is compact.
\end{proposition}

Here is the main result of this section, which relates the spectrum of
the DA chain to the spectrum of the conjugate chain.

\begin{proposition}
\label{propsamespectrum}
Assume that $|\X|=\infty$ and $|\Y|=d<\infty$. Then $K$ is a
compact operator and $\Sp(K) = \{0\} \cup\Sp(\hat{K})$.
\end{proposition}

\begin{pf}
Since $|\Y|<\infty$, $\hat{K}$ is a compact operator. It follows
from Proposition~\ref{propsamecompact} that $K$ is also compact.
Hence, $\{0\} \in\Sp(K)$, and aside from $\{0\}$, all the
elements of $\Sp(K)$ are eigenvalues of $K$. But we know from
Proposition~\ref{propsameevs} that $K$ and $\hat{K}$ share the
same eigenvalues.
\end{pf}

\begin{remark}
Liu, Wong and Kong's (\citeyear{liuwongkong1994}) Theorem 3.2 states that $\norm{K} =
\norm{\hat{K}}$ (regardless of the cardinalities of $\X$ and $\Y$).
Proposition~\ref{propsamespectrum} can be vie\-wed as a refinement
of this result in the case where $|\Y|<\infty$. See also
Roberts and Rosenthal (\citeyear{roberose2001}).
\end{remark}

In the next section we use Proposition~\ref{propsamespectrum} to
prove that the spectrum of the sandwich chain dominates the spectrum
of the DA chain.

\section{Improving the DA Algorithm}
\label{secimprove}

Suppose that $R(y,dy')$ is a Markov transition function on $\Y$ that is
reversible with respect to $f_Y(y)$. Let
$\{\tilde{X}_n\}_{n=0}^\infty$ be the sandwich chain on $\X$ whose Mtd
is given by
%
\begin{eqnarray}
\label{eqextra}
\tilde{k}(x'|x) = \int_{\Y} \int_{\Y} f_{X|Y}(x'|y')   R(y,dy') \cdot f_{Y|X}(y|x)   \nu(dy).
\end{eqnarray}
Again, routine calculations show that the sandwich chain remains
reversible with respect to the target density $f_X$. Moreover, if we
can draw from $R(y,\cdot)$, then we can draw from $\tilde{k}(\cdot|x)$
in three steps. First, draw $Y \sim f_{Y|X}(\cdot|x)$, call the
result $y$, then draw $Y' \sim R(y,\cdot)$, call the result $y'$, and
finally draw $X' \sim f_{X|Y}(\cdot|y')$.

Note that $\tilde{k}$ is not defined as the integral of the product of
two conditional densities, as in \eqref{eqMtd}. However, as we now
explain, if $R$ satisfies a certain property, called idempotence, then
$\tilde{k}$ can be re-expressed as the Mtd of a DA chain. The
transition function $R(y,dy')$ is called \textit{idempotent} if
$R^2(y,dy') = R(y,dy')$ where $R^2(y,dy') = \int_{\Y} R(y,dw)
R(w,dy')$. This property implies that, if we start the Markov chain
(defined by~$R$) at a fixed point $y$, then the distribution of the
chain after one step \textit{is the same} as the distribution after
two steps. For example, if $R(y,dy')$ does not depend on $y$, which
implies that the Markov chain is just an i.i.d. sequence, then~$R$ is
idempotent. Here is a more interesting example. Take $\Y=
\mathbb{R}$ and $R(y,dy') = r(y'|y) \, dy'$ with
\begin{eqnarray*}
r(y'|y) = e^{-|y'|}  \bigl[ I_{[0,\infty)}(y) I_{[0,\infty)}(y')
+I_{(-\infty,0)}(y) I_{(-\infty,0)}(y')  \bigr]  .
\end{eqnarray*}
It is easy to show that $\int_{\mathbb{R}} r(y'|w)   r(w|y) \, dw =
r(y'|y)$, so $R$ is indeed idempotent. Note that the chain is
reducible since, for example, if it is started on the positive
half-line, it can never get to the negative half-line. In fact,
reducibility is a common feature of idempotent chains. Fortunately,
the sandwich chain does not inherit this property.

Hobert and Marchev (\citeyear{hobemarc2008}) proved that if $R$ is
idempotent, then
%
\begin{equation}
\label{eqneed}
\tilde{k}(x'|x) = \int_{\Y} f^*_{X|Y}(x'|y)   f^*_{Y|X}(y|x)
\nu(dy)  ,
\end{equation}
where
\[
f^*(x,y) = f_Y(y) \int_\Y f_{X|Y}(x|y')   R(y, dy')  .
\]
Note that $f^*$ is a probability density (with respect to $\mu\times
\nu$) whose $x$ and $y$-marginals are $f_X$ and $f_Y$. What is
important here is not the particular form of~$f^*$, but the fact that
such a density exists, because this shows that the sandwich chain is
actually a DA chain based on the joint density $f^*(x,y)$. Therefore,
we can use the theory developed in Section~\ref{secDA} to\vspace*{1pt} analyze the
sandwich chain. Let $\tilde{K}\dvtx  L^2_0(f_X) \rightarrow L^2_0(f_X)$
denote the operator defined by the Mtd $\tilde{k}$.
Hobert and Marchev's (\citeyear{hobemarc2008}) Corollary 1 states that $\norm{\tilde{K}} \le
\norm{K}$ (see also Hobert and Rom\'an, \citeyear{hoberoma2011}).
Here is a~refinement of
that result in the case where $|\Y|<\infty$.

\begin{theorem}
\label{thmordering}
Assume that $|\X|=\infty$, $|\Y|=d<\infty$ and that $R$ is
idempotent. Then $K$ and $\tilde{K}$ are both compact operators and
each has a spectrum that consists exactly of the point $\{0\}$ and
$d-1$ eigenvalues in $[0,1)$. Furthermore, if we denote the
eigenvalues of $K$ by
\[
0 \le\lambda_{d-1} \le\lambda_{d-2} \le\cdots\le\lambda_1 < 1  ,
\]
and those of $\tilde{K}$ by
\[
0 \le\tilde{\lambda}_{d-1} \le\tilde{\lambda}_{d-2} \le\cdots
\le
\tilde{\lambda}_1 < 1  ,
\]
then $\tilde{\lambda}_i \le\lambda_i$ for each $i \in
\{1,2,\ldots,d-1\}$.
\end{theorem}

\begin{pf}
Since $R$ is idempotent, the chains defined by $k$ and $\tilde{k}$ are both DA Markov
chains. Moreover, in both cases, the conjugate chain lives on the
finite space~$\Y$, which has $d$ elements. Therefore,
Proposition~\ref{propsamespectrum} implies that $K$ and
$\tilde{K}$ are both compact and each has a spectrum consisting of
the point $\{0\}$ and $d-1$ eigenvalues in $[0,1)$. Now, Corollary
1 of Hobert and Marchev (\citeyear{hobemarc2008}) implies that $K -
\tilde{K}$ is a
positive operator. Thus, for any $g \in L^2_0(f_X)$,
\[
\frac{\langle\tilde{K} g,g \rangle}{\langle g,g \rangle} \le
\frac{\langle K g,g \rangle}{\langle g,g \rangle}  .
\]
The eigenvalue ordering now follows from an extension of the argument
used to prove Mira and Geyer's (\citeyear{mirageye1999}) Theorem 3.3. Indeed, the
Courant--Fischer--Weyl minmax characterization of eigenvalues of
compact, self-adjoint operators (see, e.g., Voss, \citeyear{voss2003}) yields
\begin{eqnarray*}
\tilde{\lambda}_i = \min_{\operatorname{dim}(V) = i-1}   \max_{g
\in V^\perp
 , g \ne0} \frac{\langle\tilde{K} g,g \rangle}{\langle g,g
\rangle}
 \le\min_{\operatorname{dim}(V) = i-1}   \max_{g \in
V^\perp , g
\ne0} \frac{\langle K g,g \rangle}{\langle g,g \rangle} = \lambda_i
 ,
\end{eqnarray*}
where $V$ denotes a subspace of $L^2_0(f_X)$ with \mbox{dimension}
$\operatorname{dim}(V)$, and $V^\perp$ is its orthogonal
complement.
\end{pf}

Theorem~\ref{thmordering} shows that, unless the two spectra are
exactly the same, $\Sp(\tilde{K})$ is closer than $\Sp(K)$ to the
ideal spectrum, $\{0\}$. In fact, in all of the numerical comparisons
that we have performed, it has always turned out that there is strict
inequality between the eigenvalues (except, of course, when they are
both zero). When the domination is strict, there exists a~positive
integer $N$ such that, for all $n \ge N$,
\begin{eqnarray*}
\int_\X\frac{ | \tilde{k}^n(x'|x) - f_X(x')  |^2}{f_X(x')}
\mu(dx')
< \int_\X\frac{ | k^n(x'|x) - f_X(x')  |^2}{f_X(x')}
  \mu(dx')  .
\end{eqnarray*}
Indeed, let $\{(\tilde{\lambda}_i,\tilde{g}_i)\}_{i=1}^{d-1}$ denote a
set of (orthonormal) eigen-solutions of $\tilde{K}$. Then, according
to \eqref{eqcs}, the $\chi^2$-distance between the distribution of
$\tilde{X}_n$ and the stationary distribution is given by
%
\begin{equation}
\label{eqics}
\sum_{i=1}^{d-1} \tilde{\lambda}_i^{2n} \tilde{g}^2_i(x)  .
\end{equation}
Now, fix $i \in\{1,\ldots,d-1\}$. If $\tilde{\lambda}_i = \lambda_i =
0$, then the $i$th term in the sum is irrelevant. On the other hand,
if $0 \le\tilde{\lambda}_i < \lambda_i$, then, no matter what the
values of $g_i(x)$ and $\tilde{g}_i(x)$ are, $\tilde{\lambda}_i^{2n}
\tilde{g}^2_i(x)$ will be less than $\lambda_i^{2n} g^2_i(x)$ for all
$n$ eventually.

In the next section we provide examples where the sandwich chain
converges \textit{much} faster than the DA chain, despite the fact
that the two are essentially equivalent in terms of computer time per
iteration.

\section{Improving the DA Algorithm for Bayesian Mixtures}
\label{secapp}

\subsection{The Model and the MDA Algorithm}
\label{secbmm}

Let $\Theta\subseteq\mathbb{R}^l$ and consider a parametric family
of densities (with respect to the Lebesgue or counting measure on
$\mathbb{R}^s$) given by $ \{h_\theta(\cdot) \dvtx  \theta\in\Theta
 \}$. We work with a $k$-component mixture of these densities
that takes the form
%
\begin{equation}
\label{eql}
f(z|\bt,\p) = \sum_{j=1}^k p_j h_{\theta_j}(z)  ,
\end{equation}
where $\bt= (\theta_1,\ldots,\theta_k)^T \in\Theta^k$ and $\p=
(p_1,\ldots, p_k)^T \in S_k$, where
\[
S_k :=  \{ \p\in\mathbb{R}^k\dvtx  p_i \in[0,1]     \mbox{ and }
p_1+\cdots+p_k=1  \}  .
\]
Let $Z_1,\ldots,Z_m$ be a random sample from $f$ and consider a
Bayesian analysis of these data. We take the prior for $\bt$ to be
$\prod_{j=1}^k \pi(\theta_j)$, where $\pi\dvtx  \Theta\rightarrow
[0,\infty)$ is a proper prior density on $\Theta$. The prior on $\p$
is taken to be the uniform distribution on $S_k$. (The results in
this section all go through with obvious minor changes if the prior on
$\p$ is taken to be symmetric Dirichlet, or if $\p$ is known and all
of its components are equal to $1/k$.) Letting $\z= (z_1,\ldots,z_m)$
denote the observed data, the posterior density is given by
%
\begin{eqnarray}
\label{eqpost}
\pi(\bt,\p|\z) = \frac{(k-1)!   I_{S_k}(\p)  [ \prod_{j=1}^k
\pi(\theta_j)  ] f(\z|\bt,\p)}{m(\z)},
\end{eqnarray}
where
\[
f(\z|\bt,\p) = \prod_{i=1}^m \Biggl [ \sum_{j=1}^k p_j
h_{\theta_j}(z_i)  \Biggr]  ,
\]
and $m(\z)$ denotes the marginal density. The complexity of this
posterior density obviously depends on many factors, including the
choices of $h_\theta$ and~$\pi$, and the observed data. However, the
versions of $\pi(\bt,\p|\z)$ that arise in practice are nearly
always
highly intractable. Moreover, as we now explain, every version of
this posterior density satisfies an interesting symmetry property,
which can render MCMC algorithms ineffectual.

The prior distribution on $(\bt,\p)$ is exchangeable in the sense
that, if $E$ is any permutation matrix of dimension $k$, then the
prior density of the point $(\bt,\p)$ is equal to that of
$(E\bt,E\p)$. Furthermore, the likelihood function satisfies a
similar invariance. Indeed, $f(\z|E\bt,E\p)$ does not vary with
$E$. Consequently, $\pi(E\bt,E\p|\z)$ is invariant to $E$, which
means that any posterior mode has $k!-1$ exact replicas somewhere else
in the space. Now, if a set of symmetric modes are separated by areas
of very low (posterior) probability, then it may take a very long time
for a~Markov chain [with invariant density $\pi(\bt,\p|\z)$] to
move from one to the other.

We now describe the MDA algorithm for exploring the mixture posterior.
Despite the fact that this algorithm has been around for many years
(Diebolt and Robert, \citeyear{diebrobe1994}), we provide a careful
description here, as this
will facilitate our development of the FS algorithm. Consider a new
(joint) density given~by
%
\begin{equation}
\label{eqaugmodel}
f(z,y|\bt,\p) = \sum_{j=1}^k p_j I_{\{j\}}(y) h_{\theta_j}(z)  .
\end{equation}
Integrating $z$ out yields the marginal mass\vspace*{1pt} function of $Y$, which is
$\sum_{j=1}^k p_j I_{\{j\}}(y)$. Hence, $Y$ is\vspace*{1pt} a~multinomial random
variable that takes the values $1,\ldots,k$ with probabilities
$p_1,\ldots,p_k$. Summing out the $y$ component leads to
%
\begin{equation}
\label{eqlatent}
\sum_{y=1}^k f(z,y|\bt,\p) = \sum_{j=1}^k p_j h_{\theta_j}(z)  ,
\end{equation}
which is just \eqref{eql}. Equation~\eqref{eqlatent} establishes
$Y$ as a~latent variable. Now suppose that $\{(Y_i,Z_i)\}_{i=1}^m$
are i.i.d. pairs from \eqref{eqaugmodel}. Their joint density is given by
\[
f(\z,{\mathbf{y}}|\bt,\p) = \prod_{i=1}^m \Biggl [ \sum_{j=1}^k p_j
I_{\{j\}}(y_i) h_{\theta_j}(z_i)  \Biggr]  ,
\]
where ${\mathbf{y}}=(y_1,\ldots,y_m)$ takes values in $\Y$, the set
of sequences
of length $m$ consisting of positive integers between 1 and $k$.
Combining $f(\z,{\mathbf{y}}|\bt,\p)$ with our prior on $(\bt,\p)$
yields the so-called complete data posterior density given~by
%
\begin{eqnarray}
\label{eqap}
\pi(\bt,\p,{\mathbf{y}}|\z)
 = \frac{(k-1)! I_{S_k}(\p)  [
\prod_{j=1}^k \pi(\theta_j)  ] f(\z,{\mathbf{y}}|\bt,\p
)}{m(\z)}  .
\end{eqnarray}
This is a valid density since, by \eqref{eqlatent},
\[
\sum_{{\mathbf{y}}\in\Y} f(\z,{\mathbf{y}}|\bt,\p) = f(\z|\bt
,\p)  ,
\]
which in turn implies that
%
\begin{equation}
\label{eqkey}
\sum_{{\mathbf{y}}\in\Y} \pi(\bt,\p,{\mathbf{y}}|\z) = \pi
(\bt,\p|\z)  .
\end{equation}
In fact, \eqref{eqkey} is the key property of the complete data
posterior density. In words, when the ${\mathbf{y}}$ coordinate is
summed out
of $\pi(\bt,\p,{\mathbf{y}}|\z)$, we are left with the target
density. Hence,
we will have a viable MDA algorithm as long as straightforward
sampling from $\pi(\bt,\p|{\mathbf{y}},\z)$ and $\pi({\mathbf
{y}}|\bt,\p,\z)$ is possible.
Note that the roles of $x$ and $y$ from Sections~\ref{secintro},
\ref{secDA} and \ref{secimprove} are being played here by $(\bt,\p)$
and ${\mathbf{y}}$, respectively.

Now consider sampling from the two conditionals. First, it follows
from \eqref{eqap} that
%
\begin{equation}
\label{eqlgee}
 \qquad \pi({\mathbf{y}}|\bt,\p,\z) = \prod_{i=1}^m  \Biggl[ \frac{\sum
_{j=1}^k p_j
I_{\{j\}}(y_i) h_{\theta_j}(z_i)}{\sum_{l=1}^k p_l
h_{\theta_l}(z_i)}  \Biggr]  .
\end{equation}
Therefore, conditional on $(\bt,\p,\z)$, the $Y_i$'s are
independent multinomial random variables and $Y_i$ takes the value
$j$
with probability $p_j h_{\theta_j}(z_i)\!/\!( \sum_{l=1}^k\! p_l
h_{\theta_l}(z_i)\!)$\vspace*{2pt} for $j \in\{1,\ldots,k\}$. Consequently,
simulating from $\pi({\mathbf{y}}| \bt,\p,\z)$ is simple.

A two-step method is used to sample from $\pi(\bt,\p| {\mathbf{y}},\z)$.
Indeed, we draw from $\pi(\p|{\mathbf{y}},\z)$ and then  from
$\pi(\bt|\p,{\mathbf{y}},\z)$. It follows from \eqref{eqap} that
\[
\pi(\p|\bt,{\mathbf{y}},\z) \propto I_{S_k}(\p) \prod_{j=1}^k
p_j^{c_j}  ,
\]
where $c_j = \sum_{i=1}^m I_{\{j\}}(y_i)$.\vspace*{2pt} This formula reveals two
facts: (i) given $(\z,{\mathbf{y}})$, $\p$ is conditionally
independent of~$\bt$, and (ii) the conditional distribution of $\p$ given
$(\z,{\mathbf{y}})$ is Dirichlet. Thus, it is easy to draw from
$\pi(\p|{\mathbf{y}},\z)$, and our sequential strategy will be
viable as long as
we can draw from $\pi(\bt|\p,{\mathbf{y}},\z)$. Our ability to
sample from
$\pi(\bt|\p,{\mathbf{y}},\z)$ will depend on the particular forms of
$h_\theta$ and the prior $\pi$. In cases where~$\pi$ is a conjugate
prior for the family $h_\theta$, it is usually straightforward to draw
from $\pi(\bt|\p,{\mathbf{y}},\z)$. For several detailed examples, see
Chapter 9 of Robert and Casella (\citeyear{robecase2004}).

The state space of the MDA chain is $\X= \Theta^k \times S_k$ and its
Mtd is given by
\[
k(\bt',\p'|\bt,\p) = \sum_{{\mathbf{y}}\in\Y} \pi(\bt',\p
'|{\mathbf{y}},\z)
\pi({\mathbf{y}}|\bt,\p,\z)  .
\]
Since $|\Y| = k^m$, Proposition~\ref{propsamespectrum} implies that
the operator $K\dvtx L^2_0  ( \pi(\bt,\p|\z)  ) \rightarrow L^2_0
 ( \pi(\bt,\p|\z)  )$ defined by $k(\bt', \p'|\bt,\p)$
is compact
and
\[
\Sp(K) = \{0,\lambda_{k^m-1},\lambda_{k^m-2},\ldots,\lambda_1\}  ,
\]
where $0 \,{\le}\,\lambda_{k^m-1} \,{\le}\,\lambda_{k^m-2}\,{\le}\,\cdots\,{\le}\,
\lambda_1 \,{<}\,1$, and the~$\lambda_i$'s are the eigenvalues of the $k^m
\times k^m$ Mtm defined by
\[
\hat{k}({\mathbf{y}}'|{\mathbf{y}}) = \int_{\Theta^k} \int_{S_k}
\pi({\mathbf{y}}'|\bt,\p,\z)
\pi(\bt,\p|{\mathbf{y}},\z) \, d\p\, d\bt .
\]
As far as we know, there are no theoretical results available
concerning the magnitude of the $\lambda_i$'s. On the other hand, as
mentioned in Section~\ref{secintro}, there is a~great deal of
empirical evidence suggesting that the MDA chain converges very
slowly because it moves between the symmetric modes of the posterior
too infrequently. In the next section we describe an alternative
chain that moves easily among the modes.

\subsection{Fr\"{u}hwirth-{S}chnatter's Algorithm}
\label{secls}

One iteration of the MDA chain can be represented graphically as
$(\bt,\p) \rightarrow{\mathbf{y}}\rightarrow(\bt',\p')$. To encourage
transitions between the symmetric modes of the posterior,
Fr\"uhwirth-Schnatter (\citeyear{fruh2001}) suggested adding an extra
step to get $(\bt,\p)
\rightarrow{\mathbf{y}}\rightarrow{\mathbf{y}}' \rightarrow(\bt
',\p')$, where the
transition ${\mathbf{y}}\rightarrow{\mathbf{y}}'$ is a random label
switching move that
proceeds as follows. Randomly choose one of the $k!$ permutations of
the integers $1,\ldots,k$, and then switch the labels in ${\mathbf
{y}}$ according
to the chosen permutation to get ${\mathbf{y}}'$. For example, suppose that
$m=8$, $k=4$, ${\mathbf{y}}=(3,3,4,1,3,3,4,3)$, and that the chosen permutation
is $(1324)$. Then we move from ${\mathbf{y}}$ to ${\mathbf
{y}}'=(2,2,1,3,2,2,1,2)$.
Using both theory and examples, we will demonstrate that
Fr\"uhwirth-Schnatter's (\citeyear{fruh2001}) Markov chain, which we call the FS chain, explores
$\pi(\bt,\p|\z)$ much more effectively than the MDA chain.\looseness=-1

To establish that the results developed in Section~\ref{secimprove}
can be used to compare the FS and MDA chains, we must show that the FS
chain is a sandwich chain with an idempotent $r$. That is, we
must\vadjust{\goodbreak}
demonstrate that the Mtd of the FS chain can be expressed in the form
%
\begin{eqnarray}
\label{eqrepresent}
\tilde{k}(\bt',\p'|\bt,\p)
 = \sum_{{\mathbf{y}}\in\Y} \sum
_{{\mathbf{y}}' \in
\Y} \pi(\bt',\p'|{\mathbf{y}}',\z)   r({\mathbf{y}}'|{\mathbf
{y}})   \pi({\mathbf{y}}|\bt,\p,\z),
\end{eqnarray}
where $r({\mathbf{y}}'|{\mathbf{y}})$ is a Mtm (on $\Y$) that is
both reversible with
respect to
\[
\pi({\mathbf{y}}|\z) = \int_{S_k} \int_{\Theta^k} \pi(\bt,\p
,{\mathbf{y}}|\z) \,
d\bt\, d\p ,
\]
and idempotent. We begin by developing a formula for $r({\mathbf
{y}}'|{\mathbf{y}})$.
Let $\mathfrak{S}_k$ denote the set (group) of permutations of the
integers $1,\ldots,k$. For $\sigma\in\mathfrak{S}_k$, let~$\sigma
{\mathbf{y}}$ represent the permuted version of ${\mathbf{y}}$. For
example, if ${\mathbf{y}}=
(3,3,4,1,3,3,4,3)$ and $\sigma= (1324)$, then $\sigma{\mathbf{y}}=
(2,2,1,3,2,2,1,2)$. The label switching move,
\mbox{${\mathbf{y}}\rightarrow{\mathbf{y}}'$},
in the FS algorithm can now be represented as follows. Choose
$\sigma$ uniformly at random from~$\mathfrak{S}_k$ and move from
${\mathbf{y}}$
to ${\mathbf{y}}' = \sigma{\mathbf{y}}$. Define the \textit{orbit}
of ${\mathbf{y}}\in\Y$~as
\[
O_{{\mathbf{y}}} =  \{ {\mathbf{y}}' \in\Y\dvtx  {\mathbf{y}}' =
\sigma{\mathbf{y}}   \mbox{ for some
$\sigma\in\mathfrak{S}_k$}  \}  .
\]
The set $O_{{\mathbf{y}}}$ simply contains all the points in $\Y$
that represent
a particular clustering (or partitioning) of the $m$ observations.
For example, the point ${\mathbf{y}}= (3,3,4,1,3, 3,4,3)$ represents the
clustering of the $m=8$ observations into the three sets:
$\{1,2,5,6,8\}$, $\{3,7\}$, $\{4\}$. And, for any $\sigma\in
\mathfrak{S}_k$, $\sigma{\mathbf{y}}$ represents that same
clustering because
all we're doing is changing the labels.

We now show that, if ${\mathbf{y}}$ is fixed and $\sigma$ is chosen
uniformly at
random from $\mathfrak{S}_k$, then the random element $\sigma{\mathbf
{y}}$ has
a uniform distribution on $O_{{\mathbf{y}}}$. Indeed, suppose that
${\mathbf{y}}$
contains $u$ distinct elements, so $u \in\{1,2,\ldots,k\}$. Then, for
any fixed ${\mathbf{y}}' \in O_{{\mathbf{y}}}$, exactly $(k-u)!$ of
the $k!$ elements in
$\mathfrak{S}_k$ satisfy $\sigma{\mathbf{y}}= {\mathbf{y}}'$. Thus,
the probability
that $\sigma{\mathbf{y}}$ equals ${\mathbf{y}}'$ is given by
$(k-u)!/k!$, which does not
depend on ${\mathbf{y}}'$. Hence, the distribution is uniform. [Note
that this
argument implies that $|O_{{\mathbf{y}}}| = k!/(k-u)!$, which can also
be shown
directly.] Therefore, we can write the Mtm~$r$ as follows:
\[
r({\mathbf{y}}'|{\mathbf{y}}) = \frac{1}{|O_{{\mathbf{y}}}|} I_{\{
O_{\mathbf{y}}\}}({\mathbf{y}}')  .
\]
Since the chain driven by $r$ cannot escape from the orbit
(clustering) in which it is started, it is reduci\-ble. (Recall from
Section~\ref{secimprove} that reducibility is a~common characteristic
of idempotent Markov chains.)

A key observation that will allow us to establish the reversibility of
$r$ is that $\pi({\mathbf{y}}|\z) = \pi(\sigma{\mathbf{y}}|\z)$
for all ${\mathbf{y}}\in\Y$ and
all $\sigma\in\mathfrak{S}_k$. Indeed,
\begin{eqnarray*}
 \pi({\mathbf{y}}|\z)  = \frac{(k-1)!}{m(\z)}  \int_{\Theta^k}
[ \pi(\theta_1)
\cdots\pi(\theta_k)  ] \cdot \Biggl\{ \int_{S_k} \prod_{i=1}^m  \Biggl[
\sum_{j=1}^k p_j I_{\{j\}}(y_i) h_{\theta_j}(z_i)  \Biggr]\, d\p \Biggr\}
\,d\bt .
\end{eqnarray*}
Let $\sigma{\mathbf{y}}= {\mathbf{y}}' = (y'_1,\ldots,y'_m)$. Now,
since $y'_i =
\sigma(j) \Leftrightarrow y_i=j$, we have
\[
\sum_{j=1}^k p_j I_{\{j\}}(y'_i) h_{\theta_j}(z_i) = \sum_{j=1}^k
p_{\sigma(j)} I_{\{j\}}(y_i) h_{\theta_{\sigma(j)}}(z_i)  .
\]
Hence,
\begin{eqnarray*}
\pi(\sigma{\mathbf{y}}|\z)
= \frac{(k-1)!}{m(\z)}  \int_{\Theta^k}  [
\pi(\theta_1) \cdots\pi(\theta_k)  ]
\cdot\Biggl \{ \int_{S_k}
\prod_{i=1}^m \Biggl [ \sum_{j=1}^k p_{\sigma(j)} I_{\{j\}}(y_i)  h_{\theta_{\sigma(j)}}(z_i)  \Biggr] \,d\p \Biggr\} \,d\bt .
\end{eqnarray*}
The fact that $\pi({\mathbf{y}}|\z) = \pi(\sigma{\mathbf{y}}|\z
)$ can now be established
through a couple of simple arguments based on symmetry.

We now demonstrate that the Mtm $r$ satisfies detailed balance with
respect to $\pi({\mathbf{y}}|\z)$; that is, we will show that, for
any ${\mathbf{y}}, {\mathbf{y}}'
\in\Y$, $r({\mathbf{y}}'|{\mathbf{y}})   \pi({\mathbf{y}}|\z) =
r({\mathbf{y}}|{\mathbf{y}}')   \pi({\mathbf{y}}'|\z)$. First,
a little thought reveals that, for any two elements ${\mathbf{y}}$ and
${\mathbf{y}}'$,
only one of two things can happen: either $O_{{\mathbf{y}}} =
O_{{\mathbf{y}}'}$ or
$O_{{\mathbf{y}}} \cap O_{{\mathbf{y}}'} = \emptyset$. If
$O_{{\mathbf{y}}} \cap O_{{\mathbf{y}}'} =
\emptyset$, then $I_{\{O_{\mathbf{y}}\}}({\mathbf{y}}') = I_{\{
O_{{\mathbf{y}}'}\}}({\mathbf{y}}) = 0$, so
$r({\mathbf{y}}'|{\mathbf{y}}) = r({\mathbf{y}}|{\mathbf{y}}') = 0$
and detailed balance is satisfied. On the
other hand, if $O_{{\mathbf{y}}}\,{=}\,O_{{\mathbf{y}}'}$, then $I_{\{
O_{\mathbf{y}}\}}({\mathbf{y}}')\!=
I_{\{O_{{\mathbf{y}}'}\}}({\mathbf{y}}) = 1$ and $1/|O_{{\mathbf
{y}}}|=1/|O_{{\mathbf{y}}'}|$, so $r({\mathbf{y}}'|{\mathbf{y}})
=r({\mathbf{y}}|{\mathbf{y}}')$, and\vadjust{\goodbreak} the common value is strictly
positive. But ${\mathbf{y}}' \in
O_{{\mathbf{y}}}$ implies that ${\mathbf{y}}' = \sigma{\mathbf{y}}$
for some $\sigma\in
\mathfrak{S}_k$. Thus, $\pi({\mathbf{y}}|\z) = \pi({\mathbf
{y}}'|\z)$, and detailed
balance holds.

Finally, it is intuitively clear that $r$ is idempotent since, if we
start the chain at ${\mathbf{y}}$, then one step results in a
uniformly chosen
point from $O_{{\mathbf{y}}}$. Obviously, the state after two steps is still
uniformly distributed  over $O_{{\mathbf{y}}}$. Here's a formal proof that
$r^2({\mathbf{y}}'|{\mathbf{y}})\,{=}\,r({\mathbf{y}}'|{\mathbf{y}})$.
For ${\mathbf{y}},{\mathbf{y}}' \in\Y$, we have
\begin{eqnarray*}
r^2({\mathbf{y}}'|{\mathbf{y}}) & =& \sum_{\w\in\Y} r({\mathbf
{y}}'|\w)   r(\w|{\mathbf{y}}) \\ 
 & =&
\sum_{\w\in\Y} \frac{1}{|O_{\w}|} I_{\{O_{\w}\}}({\mathbf{y}}')
\frac{1}{|O_{{\mathbf{y}}}|} I_{\{O_{\mathbf{y}}\}}(\w) \\ & =&
\frac{1}{|O_{{\mathbf{y}}}|}
\sum_{\w\in O_{{\mathbf{y}}}} \frac{1}{|O_{\w}|} I_{\{O_{\w}\}
}({\mathbf{y}}') \\ & =&
\frac{1}{|O_{{\mathbf{y}}}|} I_{\{O_{{\mathbf{y}}}\}}({\mathbf
{y}}') \sum_{\w\in O_{{\mathbf{y}}}}
\frac{1}{|O_{{\mathbf{y}}}|} \\ & =& r({\mathbf{y}}'|{\mathbf{y}})  ,
\end{eqnarray*}
where the fourth equality follows from the fact that $\w\in
O_{{\mathbf{y}}}
\Rightarrow O_{\w} = O_{{\mathbf{y}}}$.

We have now shown that the Mtd of the FS chain can indeed be written
in the form \eqref{eqrepresent} with an appropriate $r$ that is
reversible and idempotent. Hence, Theorem~\ref{thmordering} is
applicable and implies that the operators defined by the two chains
are both compact and each has a spectrum consisting of the
point~$\{0\}$ and $k^m-1$ eigenvalues\vspace*{1pt} in $[0,1)$. Moreover,
$\tilde{\lambda}_i \le\lambda_i$ for each\vspace*{1pt} $i \in
\{1,2,\ldots,k^m-1\}$, where $\{\tilde{\lambda}_i\}_{i=1}^{k^m-1}$ and
$\{\lambda_i\}_{i=1}^{k^m-1}$ denote the ordered eigenvalues
associated with the FS and MDA chains, respectively.


Interestingly, in the special case where $m=1$, the FS algorithm
actually produces an i.i.d. sequence from the target distribution.
Recall that $\pi({\mathbf{y}}|\z) = \pi(\sigma{\mathbf{y}}|\z)$
for all ${\mathbf{y}}\in\Y$ and
all $\sigma\in\mathfrak{S}_k$. Thus, all the points in~$O_{{\mathbf{y}}}$
share the same value of $\pi(\cdot|\z)$. When $m=1$, $\Y$ contains
only $k$ points and they all exist in the same orbit. Thus,
$\pi({\mathbf{y}}|\z) = 1/k$ for all ${\mathbf{y}}\in\Y$.
Moreover, since there is only
one orbit, $r({\mathbf{y}}'|{\mathbf{y}})=1/k$ for all ${\mathbf
{y}}' \in\Y$, that is, the Markov
chain corresponding to $r$ is just an i.i.d. sequence from the uniform
distribution on $\Y$. In other words, the label switching move
results in an exact draw from $\pi({\mathbf{y}}'|\z)$. Now recall
the graphical
representation of one iteration of the FS algorithm: $(\bt,\p)
\rightarrow{\mathbf{y}}\rightarrow{\mathbf{y}}' \rightarrow(\bt
',\p')$. When
$m=1$, the arguments above imply that, given $(\bt,\p)$, the
density of $({\mathbf{y}},{\mathbf{y}}',\bt',\p')$ is
\begin{eqnarray*}
\pi({\mathbf{y}}|\bt,\p,\z) r({\mathbf{y}}'|{\mathbf{y}}) \pi
(\bt',\p'|{\mathbf{y}}',\z)
 =\pi({\mathbf{y}}|\bt,\p,\z) \pi({\mathbf{y}}'|\z) \pi(\bt',\p
'|{\mathbf{y}}',\z)  .
\end{eqnarray*}
Thus, conditional on $(\bt,\p)$, ${\mathbf{y}}$ and $({\mathbf
{y}}',\bt',\p')$ are
independent, and the latter has density
\[
\pi({\mathbf{y}}'|\z) \pi(\bt',\p'|{\mathbf{y}}',\z) = \pi(\bt
',\p',{\mathbf{y}}'|\z)  .
\]
It follows that, marginally, $(\bt',\p') \sim\pi(\bt',\p'|\z)$, so
the FS algorithm produces an i.i.d.  sequence from the target posterior
density. When $m=1$, $|\Y| = k^m = k$. Thus, while the spectrum of
the MDA operator contains $k-1$ eigenvalues, at least one of which is
strictly positive, the spectrum of the FS operator is the ideal
spectrum, $\{0\}$.

In the next section we consider two specific mixture models and, for
each one, we compare the spectra associated with FS and MDA chains.
The first example is a toy problem where we are able to get exact
formulas for the eigenvalues. The second example is a normal mixture
model that is frequently used in practice, and we approximate the
eigenvalues via classical Monte Carlo methods.

\section{Examples}
\label{secexamples}

\subsection{A Toy Bernoulli Mixture}
\label{sectoy}

Take the parametric family $h_\theta$ to be the family~of Bernoulli
mass functions, and consider a two-component version of the mixture
with known weights both equal to $1/2$. This mixture density takes
the form
\[
f(z|r,s) = \tfrac{1}{2} r^{z} (1-r)^{1-z} + \tfrac{1}{2} s^{z}
(1-s)^{1-z}  ,
\]
where $z \in\{0,1\}$ and $\bt=(r,s)$. To simplify things ever
further, assume that $r,s \in\{\rho,1-\rho\}$ where $\rho\in
(0,1/2)$ is fixed; that is, the two success probabilities, $r$ and
$s$, can only take the values $\rho$ and $1-\rho$. Hence, $(r,s) \in
\X=  \{ (\rho,\rho), (\rho,1-\rho), (1-\rho,\rho), (\mbox{$1-\rho$}, 1-\rho) \}$.
Our prior for $(r,s)$ puts mass $1/4$ on each of these four
points. A simple calculation shows that the posterior mass function
takes the form
\begin{eqnarray*}
\pi(r,s|\z) =
\frac{I_{\{\rho,1-\rho\}}(r) I_{\{\rho,1-\rho\}}(s) (r+s)^{m_1} (2-r-s)^{m-m_1}}{2^m \rho^{m_1}
(1-\rho)^{m-m_1} + 2^m \rho^{m-m_1} (1-\rho)^{m_1} + 2}  ,
\end{eqnarray*}
where $\z=(z_1,\ldots,z_m) \in\{0,1\}^m$ denotes the obser\-ved data,
and $m_1$ denotes the number of successes among the $m$ Bernoulli
trials, that is, $m_1 = \sum_{i=1}^m z_i$. While we would never
actually use MCMC to explore this simple four-point posterior, it is
both interesting and useful to compare the FS and MDA algorithms in
this context.


As described in Section~\ref{secbmm}, the MDA algorithm is based
on the complete data posterior density, which is denoted here by
$\pi(r,s,{\mathbf{y}}|\z)$. (The fact that $\p$ is known in this
case doesn't
really change anything.) Of course, all we really need are the
specific forms of the conditional mass functions, $\pi({\mathbf
{y}}|r,s,\z)$ and
$\pi(r,s|{\mathbf{y}},\z)$. It follows from the general development in
Section~\ref{secbmm} that, given $(r,s,\z)$, the components of
${\mathbf{y}}=(y_1,y_2,\ldots,y_m)$ are independent multinomials with mass
functions given by
\begin{eqnarray*}
\pi(y_i|r,s,\z)
= \frac{I_{\{1\}}(y_i) r^{z_i} (1-r)^{1-z_i} +
I_{\{2\}}(y_i) s^{z_i} (1-s)^{1-z_i}}{r^{z_i} (1-r)^{1-z_i} +
s^{z_i} (1-s)^{1-z_i}}  .
\end{eqnarray*}
Furthermore, it is easy to show that, given $({\mathbf{y}},\z)$, $r$
and $s$ are
independent so $\pi(r,s|{\mathbf{y}},\z) = \pi(r|{\mathbf{y}},\z)
  \pi(s|{\mathbf{y}},\z)$. Now,
for $j \in\{1,2\}$ and $k \in\{0,1\}$, let $m_{jk}$ denote the
number of $(y_i,z_i)$ pairs that take the value $(j,k)$. (Note that
$m_{10} + m_{11} = c_1$ and $m_{11} + m_{21} = m_1$.) Then we have
{\fontsize{10pt}{12pt}\selectfont{%
\begin{eqnarray*}
\pi(r|{\mathbf{y}},\z)
= \frac{I_{\{\rho\}}(r) \rho^{m_{11}}
(1-\rho)^{m_{10}}+
I_{\{1-\rho\}}(r) \rho^{m_{10}} (1-\rho)^{m_{11}}}{\rho^{m_{11}}
(1-\rho)^{m_{10}} + \rho^{m_{10}} (1-\rho)^{m_{11}}}  ,
\end{eqnarray*}}}%
and
{\fontsize{10pt}{12pt}\selectfont{%
\begin{eqnarray*}
\pi(s|{\mathbf{y}},\z)
= \frac{I_{\{\rho\}}(s) \rho^{m_{21}}
(1-\rho)^{m_{20}} +
I_{\{1-\rho\}}(s) \rho^{m_{20}} (1-\rho)^{m_{21}}}
{\rho^{m_{21}}
(1-\rho)^{m_{20}} + \rho^{m_{20}} (1-\rho)^{m_{21}}} .
\end{eqnarray*}}}%
The state space of the MDA chain is $\X=  \{ (\rho,\rho),
(\rho, 1-\rho), (1-\rho,\rho), (1-\rho,1-\rho)  \}$, which has only
four points. Hence, in this toy Bernoulli example, we can analyze the
MDA chain directly. Its Mtm is $4 \times4$ and the transition
probabilities are given by
%
\begin{equation}
\label{eqmtmb}
 \quad k(r',s'|r,s) = \sum_{{\mathbf{y}}\in\Y} \pi(r',s'|{\mathbf{y}},\z
)   \pi({\mathbf{y}}|r,s,\z)
 ,
\end{equation}
where $\Y= \{1,2\}^m$. We now perform an eigen-analysis of this Mtm.
Note that $\pi(r',s'|{\mathbf{y}},\z)$ and $\pi({\mathbf
{y}}|r,s,\z)$ depend on ${\mathbf{y}}$ only
through $m_{10}$, $m_{11}$, $m_{20}$ and $m_{21}$. If we let
$m_0=m-m_1$, then we can express the transition probabilities as
follows:
\begin{eqnarray*}
 k(r',s'|r,s)  &=& \sum_{i=0}^{m_1} \sum_{j=0}^{m_0} \binom{m_1}{i}
\binom{m_0}{j}
\biggl [ \frac{I_{\{\rho\}}(r') \rho^i (1-\rho)^j +
I_{\{1-\rho\}}(r') \rho^j (1-\rho)^i}{\rho^i (1-\rho)^j + \rho^j
(1-\rho)^i}  \biggr]  \\
&&  {}\cdot \biggl[ \frac{I_{\{\rho\}}(s')
\rho^{m_1-i} (1-\rho)^{m_0-j}  + I_{\{1-\rho\}}(s') \rho^{m_0-j}
(1-\rho)^{m_1-i}}{\rho^{m_1-i} (1-\rho)^{m_0-j}  + \rho^{m_0-j}
(1-\rho)^{m_1-i}}
  \biggr] \frac{r^i (1-r)^j s^{m_1-i}
(1-s)^{m_0-j}}{(r+s)^{m_1} (2-r-s)^{m_0}}  .
\end{eqnarray*}
Now, for $k=0,1,2$ define
\begin{eqnarray*}
  w_k(\rho)
 = \sum_{i=0}^{m_1} \sum_{j=0}^{m_0} \binom{m_1}{i}
\binom{m_0}{j}
 \biggl[ \frac{\rho^{k(m_0-j+i)} (1-\rho)^{k(m_1-i+j)}}{ ( \rho^i (1-\rho)^j + \rho^j (1-\rho)^i  )  (
\rho^{m_1-i} (1-\rho)^{m_0-j} + \rho^{m_0-j} (1-\rho)^{m_1-i}
 ) } \biggr]  .
\end{eqnarray*}
Using this notation, we can write the Mtm as follows:
\[
k = \left [
\matrix{ \frac{\rho^{m_1} (1-\rho)^{m_0}}{2^m}w_0(\rho) & \frac{1}{2^m}
w_1(\rho)& \frac{1}{2^m} w_1(\rho) &\frac{\rho^{m_0} (1-\rho)^{m_1}}{2^m} w_0(\rho) \vspace*{2pt}\cr
\rho^{m_1} (1-\rho)^{m_0} w_1(\rho) & w_2(\rho)& \rho^m(1-\rho)^m w_0(\rho) & \rho^{m_0} (1-\rho)^{m_1} w_1(\rho) \vspace*{2pt}\cr
\rho^{m_1} (1-\rho)^{m_0} w_1(\rho) & \rho^m (1-\rho)^m w_0(\rho)& w_2(\rho) & \rho^{m_0} (1-\rho)^{m_1} w_1(\rho) \vspace*{2pt}\cr
\frac{\rho^{m_1} (1-\rho)^{m_0}}{2^m} w_0(\rho) & \frac{1}{2^m}w_1(\rho)
& \frac{1}{2^m} w_1(\rho) &\frac{\rho^{m_0} (1-\rho)^{m_1}}{2^m} w_0(\rho)
 }\right]  .
\]
We have ordered the points in the state space as follows:
$(\rho,\rho)$, $(\rho,1-\rho)$, $(1-\rho,\rho)$ and
$(1-\rho,1-\rho)$. So, for example, the element in the second row,
third column is the probability of moving from $(\rho,1-\rho)$ to
$(1-\rho,\rho)$. Note that all of the transition probabilities are
strictly positive, which implies that the MDA chain is Harris ergodic.

Of course, since $k$ is a Mtm, it satisfies $k v_0 = \lambda_0 v_0$
where $v_0 = \mathbf{1}$ and $\lambda_0 = 1$. Again, $(v_0,\lambda_0)$
does not count as an eigen-solution for us because we are using
$L^2_0(f_X)$ instead of $L^2(f_X)$, and the only constant function in
$L^2_0(f_X)$ is 0. For us, there are three eigen-solutions, and we
write them as $(v_i,\lambda_i)$, $i \in\{1,2,3\}$, where $0 \le
\lambda_3 \le\lambda_2 \le\lambda_1 < 1$. Note that the first and
fourth rows of $k$ are identical, which means that $\lambda_3=0$. The
remaining eigen-solutions follow from the general results in the
\hyperref[appm]{Appendix}. Indeed,
\[
\lambda_1 = w_2(\rho) - \rho^m (1-\rho)^m w_0(\rho)  ,
\]
and the corresponding eigen-vector is $v_1 = (0, 1, -1,  0)^T$.
Finally,
\[
\lambda_2 = \frac{g(\rho)w_0(\rho)}{2^m} - g(\rho)w_1(\rho)
\]
and $v_2 = (\alpha, 1, 1, \alpha)^T$, where $g(\rho) = \rho^{m_1}
(1-\rho)^{m_0} + \rho^{m_0} (1-\rho)^{m_1}$ and
\[
\alpha= \frac{g(\rho) w_0(\rho) - 2^m}{2^m g(\rho) w_1(\rho)}  .
\]
(The fact that $\lambda_2 \le\lambda_1$ actually follows from our
analysis of the FS chain below.) We now use these results to
demonstrate that the MDA algorithm can perform quite poorly for the
Bernoulli model.

\begin{figure*}

\includegraphics{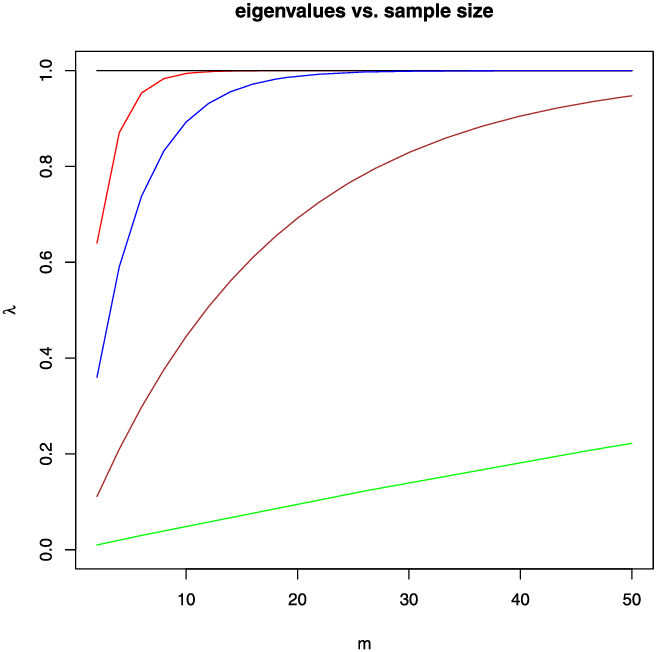}

\caption{The behavior of the dominant eigenvalue for the MDA chain in
the Bernoulli model. The graph shows how the dominant eigenvalue of
the MDA chain changes with sample size, $m$, for several different
values of $\rho$, in the case where half the $z_i$'s are 0 and the
other half are 1. (Only even sample sizes are considered.) The
red, blue, brown and green lines correspond to $\rho$ values of
$1/10$, $1/5$, $1/3$ and $9/20$, respectively.}
\label{figfigureev}
\end{figure*}

Consider a numerical example in which $m=10$, $\rho=1/10$ and the data
are $z_1=\cdots=z_5=0$ and $z_6=\cdots=z_{10}=1$. The posterior mass
function is as follows:
\[
\pi(\rho,\rho|\z) = \pi(1-\rho,1-\rho|\z) = 0.003
\]
 and
 \[
   \pi(\rho,1-\rho|\z) =
\pi(1-\rho,\rho|\z) = 0.497  .
\]
So there are two points with exactly the same very high probability,
and two points with exactly the same very low probability. The MDA
chain converges slowly due to its inability to move between the two
high probability points. Indeed, the Markov transition matrix in this
case is as follows:
\[
k =  \left[
\matrix{
0.10138 & 0.39862 & 0.39862 & 0.10138 \cr
0.00241 & 0.99457 & 0.00061 & 0.00241 \cr
0.00241 & 0.00061 & 0.99457 & 0.00241 \cr
0.10138 & 0.39862 & 0.39862 & 0.10138
}
 \right]  .
\]
Suppose we start the chain in the state $(\rho,1-\rho)$. The expected
number of steps before it reaches the other high probability state,
$(1-\rho,\rho)$, is quite large. First, we expect the chain to remain
in the state $(\rho,1-\rho)$ for about $1/(1-0.99457) \approx184$
iterations. Then, conditional on the chain leaving $(\rho,1-\rho)$,
the probability that it moves to $(\rho,\rho)$ or $(1-\rho,1-\rho)$ is
about 0.89. And if it does reach $(\rho,\rho)$ or $(1-\rho,1-\rho)$,
there is still about a 40\% chance that it will jump right back to the
point $(\rho,1-\rho)$, where it will stay for (approximately) another
184 iterations. All of this translates into slow convergence. In
fact, the two nonzero eigenvalues are $(\lambda_1,\lambda_2) =
(0.99395,0.19795)$. Moreover, the problem gets worse as the sample
size increases. For example, if we increase the sample size to $m=20$
(and maintain the $50\dvtx 50$ split of 0's and 1's in the data), then
$(\lambda_1,\lambda_2) = (0.99996,0.15195)$.
Figure~\ref{figfigureev} shows how the dominant eigenvalue,
$\lambda_1$, changes with sample size for several different values of
$\rho$. We conclude that, for fixed $\rho$, the convergence rate
deteriorates as the sample size increases. Moreover, the (negative)
impact of increasing sample size is magnified as $\rho$ gets smaller.

%

\begin{figure*}

\includegraphics{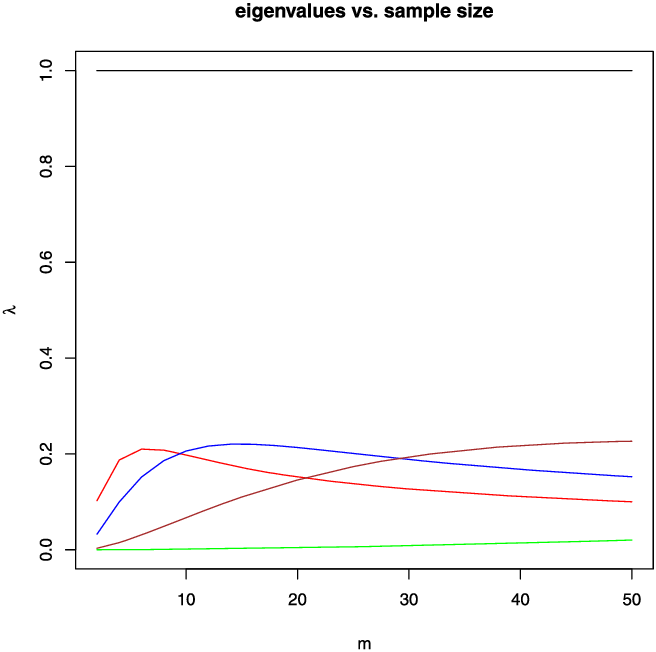}

\caption{The behavior of the dominant eigenvalue for the FS chain in
the Bernoulli model. The graph shows how the dominant eigenvalue of
the FS chain changes with sample size, $m$, for several different
values of $\rho$, in the case where half the $z_i$'s are 0 and the
other half are 1. (Only even sample sizes are considered.) The
red, blue, brown and green lines correspond to $\rho$ values of
$1/10$, $1/5$, $1/3$ and $9/20$, respectively.}
\label{figfigureevS}
\end{figure*}

Now consider implementing the FS algorithm for the Bernoulli mixture.
Because the mixture has only two components, the random label
switching step,  ${\mathbf{y}}\rightarrow{\mathbf{y}}'$, is quite
simple. Indeed, we
simply flip a fair coin. If the result is heads, then we take ${\mathbf
{y}}' = {\mathbf{y}}$, and if the result is tails, then we take ${\mathbf
{y}}'=\overline{{\mathbf{y}}}$,
where $\overline{{\mathbf{y}}}$ denotes ${\mathbf{y}}$ with its 1's
and 2's flipped. The Mtm of the FS chain has entries given by
\begin{eqnarray*}
\tilde{k}(r',s'|r,s) =\frac{1}{2} \sum_{{\mathbf{y}}\in\Y}
\pi(r',s'|\overline{{\mathbf{y}}},\z)   \pi({\mathbf{y}}|r,s,\z
)+ \frac{1}{2} \sum_{{\mathbf{y}}
\in\Y} \pi(r',s'|{\mathbf{y}},\z)   \pi({\mathbf{y}}|r,s,\z)  .
\end{eqnarray*}
It follows that
\[\tilde{k} =  \left[
\matrix{ \frac{\rho^{m_1}
(1-\rho)^{m_0}}{2^m} w_0(\rho) & \frac{1}{2^m} w_1(\rho)  &\frac{1}{2^m} w_1(\rho)   &\frac{\rho^{m_0} (1-\rho)^{m_1}}{2^m} w_0(\rho)  \vspace*{2pt}\cr
\rho^{m_1} (1-\rho)^{m_0} w_1(\rho) & \frac{w_2(\rho) +\rho^m (1-\rho)^m w_0(\rho)}{2} & \frac{w_2(\rho) +\rho^m (1-\rho)^m w_0(\rho)}{2} & \rho^{m_0}(1-\rho)^{m_1} w_1(\rho)  \vspace*{2pt}\cr
 \rho^{m_1} (1-\rho)^{m_0}w_1(\rho) & \frac{w_2(\rho) + \rho^m (1-\rho)^m w_0(\rho)}{2} & \frac{w_2(\rho) + \rho^m (1-\rho)^m w_0(\rho)}{2} & \rho^{m_0} (1-\rho)^{m_1} w_1(\rho)  \vspace*{2pt}\cr
\frac{\rho^{m_1} (1-\rho)^{m_0}}{2^m} w_0(\rho) &\frac{1}{2^m} w_1(\rho)
& \frac{1}{2^m} w_1(\rho) &\frac{\rho^{m_0} (1-\rho)^{m_1}}{2^m} w_0(\rho)
} \right]  .\]
%
%
Note that this matrix differs from $k$ only in the middle four
elements. Indeed, the $(2,2)$ and $(2,3)$ elements in $k$ have both
been replaced by their average in $\tilde{k}$, and the same is true of
the $(3,2)$ and $(3,3)$ elements. The matrix $\tilde{k}$ has rank at
most two, so there is at most one nonzero eigenvalue to find. Using
the results in the \hyperref[appm]{Appendix}\vadjust{\goodbreak} along with the eigen-analysis of $k$
performed earlier, it is easy to see that the nontrivial
eigen-solution of $\tilde{k}$ is $(\tilde{v}_1,\tilde{\lambda}_1) =
(v_2,\lambda_2)$. So, the effect on the spectrum of adding the random
label switching step is to replace the dominant eigenvalue with 0!
(Note
that Theorem~\ref{thmordering} implies that $\lambda_2 =
\tilde{\lambda}_1 \le\lambda_1$, which justifies our ordering of the
eigenvalues of $k$.) Consider again the simple numerical example with
the $50\dvtx 50$ split of 0's and 1's. In the case $m=10$, the result of
adding the extra step is to replace the dominant eigenvalue,
$0.99395$, by $0.19795$. When $m=20$, $0.99996$ is replaced by
$0.15195$. This suggests that, in contrast to the MDA algorithm,
increasing sample size does not adversely affect the FS algorithm.
More evidence for this is provided in Figure~\ref{figfigureevS},
which is the analogue of Figure~\ref{figfigureev} for the FS
algorithm. Note that the dominant eigenvalues are now substantially
smaller, and no longer converge to 1 as the sample size increases. In
fact, based on experimental evidence, it appears that, for a fixed
value of $\rho$, $\lambda_2$ hits a~maximum and then decreases with
sample size. It is surprising that such a minor change in the MDA
algorithm could result in such a huge improvement. In the next
section\vspace*{-1pt} we consider a mixture of normal
densities.\looseness=-1
%

\subsection{The Normal Mixture}
\label{secnormal}

Assume that $Z_1,\ldots,Z_m$ are i.i.d. from the density
\begin{eqnarray*}
f(z|\mu,\tau^2,p)
= p \frac{1}{\tau_1} \phi\biggl (
\frac{z-\mu_1}{\tau_1}  \biggr) + (1-p)\frac{1}{\tau_2} \phi\biggl (
\frac{z-\mu_2}{\tau_2}  \biggr)  ,
\end{eqnarray*}
where $p \in[0,1]$, $\mu= (\mu_1,\mu_2) \in\mathbb{R}^2$, $\tau^2
= (\tau^2_1,\tau^2_2) \in\mathbb{R}^2_+$, and $\phi(\cdot)$
denotes the standard normal density function. The prior for $p$ is
$\operatorname{Uniform}(0,1)$, and the prior for $(\mu,\tau^2)$
takes the
form $\pi(\mu_1,\tau^2_1)   \pi(\mu_2,\tau^2_2)$. As for $\pi$,
we use the standard (conditionally conjugate) prior given by
\[
\pi(\mu_1,\tau^2_1) = \pi(\mu_1|\tau^2_1) \pi(\tau^2_1)  ,
\]
where $\pi(\mu_1|\tau^2_1) = \mathrm{N}(0,\tau^2_1)$ and $\pi(\tau^2_1)
= \operatorname{IG}(2,1/2)$ (Robert and Casella, \citeyear
{robecase2004}, Section 9.1). By $W \sim
\operatorname{IG}(\alpha,\gamma)$, we mean that $W$ is a random
variable with
density function proportional to $w^{-\alpha-1} \exp\{-\gamma/w\}\cdot
I_{\mathbb{R_+}}(w)$. In contrast with the Bernoulli example from the
previous subsection, the posterior density associated with the normal
mixture is quite intractable and has a complicated (and uncountable)
support given by $\X= \mathbb{R}^2 \times\mathbb{R}^2_+ \times
[0,1]$.

The MDA algorithm is based on the complete-data posterior density,
which we denote here by $\pi(\mu,\tau^2,p,\allowbreak {\mathbf{y}}|\z)$.
Again, the
development in Section~\ref{secbmm} implies that, given
$(\mu,\tau^2,p,\z)$, the elements of ${\mathbf{y}}$ are independent
multinomials
and the probability that the $i$th coordinate equals 1 (which is one
minus the probability that it equals 2) is given by
%
\begin{eqnarray}
\label{eqmps}
\biggl(p \frac{1}{\tau_1} \phi \biggl( \frac{z_i-\mu_1}{\tau_1}\biggr)\biggr)
\Big/\biggl(p \frac{1}{\tau_1} \phi \biggl( \frac{z_i-\mu_1}{\tau_1}
\biggr) + (1-p) \frac{1}{\tau_2} \phi \biggl(\frac{z_i-\mu_2}{\tau_2}  \biggr)\biggr).
\end{eqnarray}
We sample $\pi(\mu,\tau^2,p|{\mathbf{y}},\z)$ via sequential
sampling from
$\pi(p|{\mathbf{y}},\z)$ and $\pi(\mu,\tau^2|p,{\mathbf{y}},\z
)$. The results in
Section~\ref{secbmm} show that $p|{\mathbf{y}},\z\sim
\operatorname{Beta}(c_1+1,c_2+1)$. Moreover, it's easy to show that, given
$(p,{\mathbf{y}},\z)$, $(\mu_1, \tau^2_1)$ and $(\mu_2,\tau^2_2)$ are
independent. Routine calculations show that
\[
\mu_1|\tau_1^2,p,{\mathbf{y}},\z\sim\mathrm{N}  \biggl( \frac{c_1}{c_1+1}
\overline{z}_1, \frac{\tau^2_1}{(c_1+1)}  \biggr)
\]
and
\[
\tau_1^2 |p,{\mathbf{y}},\z\sim\operatorname{IG} \biggl ( \frac
{c_1+4}{2}, \frac{1}{2}
 \biggl( s_1^2 + \frac{c_1 \overline{z}_1^2}{(c_1+1)} + 1  \biggr)  \biggr)
 ,
\]
where $\overline{z}_1 = \frac{1}{c_1} \sum_{i=1}^m I_{\{ 1 \}}(y_i)
z_i$ and $s_1^2 = \sum_{i=1}^m I_{\{ 1 \}}(y_i)\cdot  (z_i -
\overline{z}_1)^2$. Of course, the distribution of
$(\mu_2,\tau^2_2)$ given $(p,{\mathbf{y}},\z)$ has an analogous form.

%
\begin{figure*}

\includegraphics{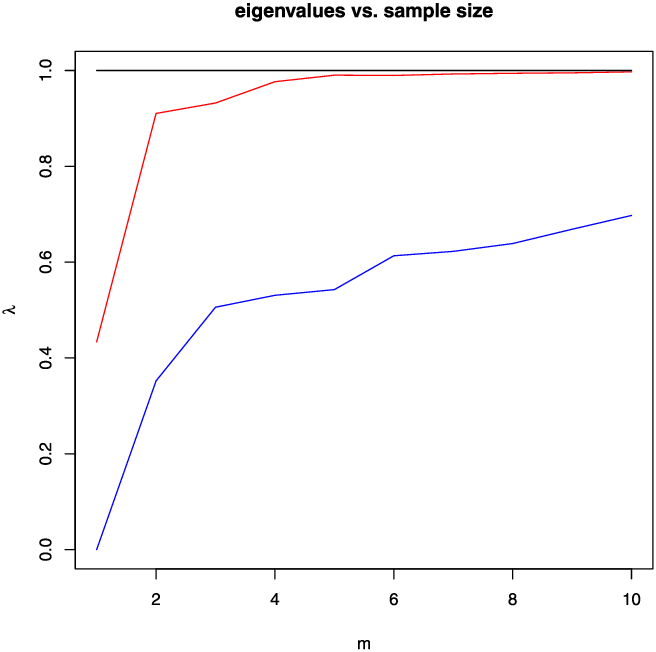}

\caption{The behavior of the dominant eigenvalue for the MDA and FS
chains in the normal model. The graph is based on the first
simulated data set and shows how the dominant eigenvalue changes
with sample size, $m$, for the MDA algorithm (red line) and the FS
algorithm (blue line).}
\label{fignorm1}
\end{figure*}

The results developed in Section~\ref{secDA} imply that the spectrum
of the operator associated with the MDA chain consists of the point
$\{0\}$ and the eigenvalues of the Mtm of the conjugate chain, which
lives on $\Y= \{1,2\}^m$. Unfortunately, the Mtm of the conjugate
chain is also intractable. Indeed, a generic element of this matrix
has the following form:
\begin{eqnarray*}
\hat{k}({\mathbf{y}}'|{\mathbf{y}})
= \int_0^1 \int_{\mathbb
{R}^2_+} \int_{\mathbb{R}^2}
\pi({\mathbf{y}}'|\mu,\tau^2,p,\z)
\cdot  \pi(\mu,\tau^2,p|{\mathbf
{y}},\z) \, d\mu\,
d\tau^2 \, dp.
\end{eqnarray*}
This integral cannot be computed in closed form. In particular,
$\pi({\mathbf{y}}'|\mu,\tau^2,p,\z)$ is the product of $m$
probabilities of the
form \eqref{eqmps}, and the sums in the denominators of these
probabilities render the integral intractable. However, note that
$\hat{k}({\mathbf{y}}'|{\mathbf{y}})$ can be interpreted as the
expected value of
$\pi({\mathbf{y}}'|\mu,\tau^2,p,\z)$ with respect to the density
$\pi(\mu,\tau^2,p|{\mathbf{y}},\z)$. Of course, for fixed $\z$,
we know how to
draw from $\pi(\mu,\tau^2,p|{\mathbf{y}},\z)$, and we have
$\pi({\mathbf{y}}'|\mu,\tau^2,p,\z)$ in closed form. We therefore
have the
ability to estimate $\hat{k}({\mathbf{y}}'|{\mathbf{y}})$ using
classical Monte Carlo.
Once we have an estimate of the entire $2^m \times2^m$ Mtm, we can
calculate its eigenvalues.

The same idea can be used to approximate the eigenvalues of the FS
chain. The results in Section~\ref{secimprove} show that we can
express the FS algorithm as a DA algorithm with respect to an
alternative complete-data posterior density, which we write as
$\pi^*(\mu,\tau^2,p,{\mathbf{y}}|\z)$. The eigenvalues of the operator
defined by the FS chain are the same as those of the Mtm in which the
probability of the transition ${\mathbf{y}}\rightarrow{\mathbf{y}}'$
is given by
\begin{eqnarray*}
\int_0^1 \int_{\mathbb{R}^2_+} \int_{\mathbb{R}^2}\pi^*({\mathbf{y}}'|\mu,\tau^2,p,\z)
\cdot\pi^*(\mu,\tau^2,p|{\mathbf{y}},\z) \, d\mu\,d\tau^2 \, dp  .
\end{eqnarray*}
It is straightforward to simulate from $\pi^*(\mu,\tau^2,p|{\mathbf
{y}},\z)$,
and $\pi^*({\mathbf{y}}'|\mu,\tau^2,p,\z)$ is available in closed form.

To use our classical Monte Carlo idea to estimate the spectra
associated with the MDA and FS chains, we must specify the data, $\z$.
Furthermore, the Bernoulli example in the previous subsection showed
that the convergence rates of the two algorithms can depend heavily on
the sample size, $m$. Thus, we would like to \mbox{explore} how an
increasing sample size affects the convergence rates of the MDA and FS
chains in the current context. To generate data, we simulated a
random sample of size 10 from a $50\dvtx 50$ mixture of a $\mathrm{N}(0,0.55^2)$
and a $\mathrm{N}(3,0.55^2)$, and this resulted in the following
observations:
\begin{eqnarray*}
\z = (z_1,\ldots,z_{10})
 =(0.2519, 2.529, -0.2930, 2.799,
3.397,
0.5596, 2.810, 2.541, 2.487, -0.1937)  .
\end{eqnarray*}
We considered 10 different data sets ranging in size from $m=1$ to
$m=10$. The first data set contained the single point $z_1 =
0.25192$, the second contained the first two observations $(z_1,z_2) =
(0.25192, 2.5287)$, the third contained $(z_1,z_2,z_3) = (0.25192,
2.5287,  -0.29303)$, and so on up to the tenth data set, which
contained all ten observations. For each of these 10 data sets, we
used the classical Monte Carlo technique described above to estimate
the Mtm for both the MDA and FS algorithms. In particular, for each
row of the Mtm we used a single Monte Carlo sample of size 200,000
[from $\pi(\mu,\tau^2,p|{\mathbf{y}},\z)$ for DA, and from
$\pi^*(\mu,\tau^2,p|{\mathbf{y}},\z)$ for FS] to estimate each of
the entries in
that row. We then calculated the eigenvalues of the estimated Mtms
and recorded the largest one. The results are shown in
Figure~\ref{fignorm1}, which has some interesting features. Note
that the dominant eigenvalues of the MDA chain are much closer to 1 than
the corresponding dominant eigenvalues of the FS chain. Even at
$m=5$, the dominant eigenvalue of the MDA chain is already above 0.99.
As in the previous example, the convergence rate of the MDA chain
deteriorates as $m$ increases. It is not clear whether the FS chain
slows down as $m$ increases. It may be the case that the FS
eigenvalue would eventually level off, or perhaps the FS chain would
eventually begin to speed up, as in the Bernoulli example. Note that,
as proven in Section~\ref{secls}, when $m=1$, the FS eigenvalue
is 0. (To ascertain the accuracy of our estimates, we repeated the
entire classical Monte Carlo simulation 6 times, with different random
number seeds, and based on this, we believe that our eigenvalue
estimates are correct up to three decimal places.)

In the case where all 10 observations are considered, the dimension of
the Mtms is $1024 \times1024$, and each element must be estimated by
classical Monte Carlo. Thus, while it would be very interesting to
consider larger sample sizes (beyond 10), and even mixtures with more
than 2 components, the matrices become quite unwieldy.

\begin{figure*}

\includegraphics{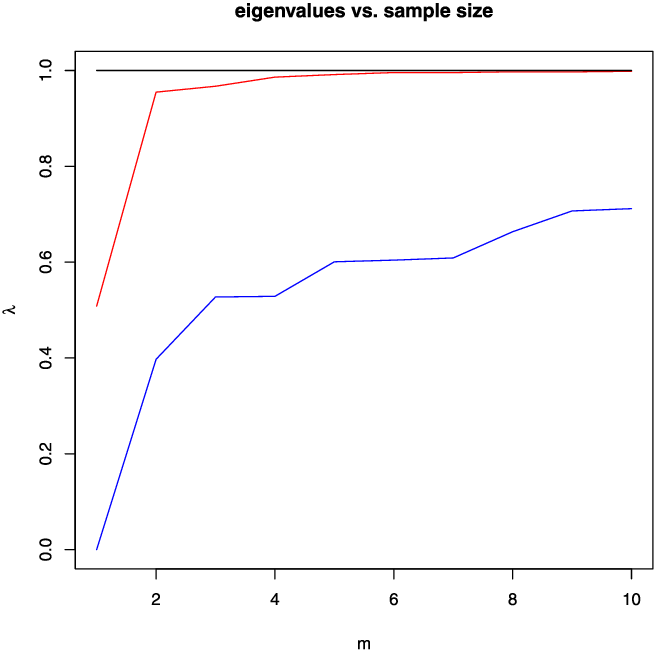}
\caption{The behavior of the dominant eigenvalue for the MDA and FS
chains in the normal model. The graph is based on the second
simulated data set and shows how the dominant eigenvalue changes
with sample size, $m$, for the MDA algorithm (red line) and the FS
algorithm (blue line).}
\label{fignorm2}
\end{figure*}

We simulated a second set of 10 observations from the same $50\dvtx 50$
mixture and repeated the entire process for the purpose of validation.
The second simulation resulted in the following data:
\begin{eqnarray*}
\z= (z_1,\ldots,z_{10})=(0.6699, 3.408, 0.1093, 3.289,-0.1407,3.525, 2.454, 0.2716, -0.7443, 3.570).
\end{eqnarray*}
Figure~\ref{fignorm2} is the analogue of Figure~\ref{fignorm1} for
the second simulation. The results are nearly identical to those from
the first simulation.

%

\setcounter{equation}{0}
\begin{appendix}\label{appm}
\label{secappendix}

\section*{Appendix}

Consider a Mtm of the form
\[
M =  \left[
\matrix{ a & b & b & c \cr
d & e & f & \frac{cd}{a} \vspace*{2pt}\cr
d & f & e & \frac{cd}{a} \cr
a & b & b & c
}
 \right]  ,
\]
and assume that all of the elements are strictly positive, so the
corresponding Markov chain is irreducible and aperiodic. Note that
both of the Mtms studied in Section~\ref{sectoy} have this form.
Routine manipulation shows that $M$ is reversible with respect to
$(\pi_1, \pi_2, \pi_3, \pi_4)^T$ where $\pi_1=ad/(ad+2ab+cd)$,
$\pi_2=b\pi_1/d$, $\pi_3=\pi_2$ and $\pi_4=c\pi_1/a$. In the
remainder of this section we perform an eigen-analysis of the matrix $M$.

Of course, since $M$ is a Mtm, it satisfies $k v_0 = \lambda_0 v_0$
where $v_0 = \mathbf{1}$ and $\lambda_0 = 1$. Furthermore, since the
first and fourth rows are equal, there is at least one eigenvalue
equal to zero. Indeed, $M v_3 = 0$, where $v_3 = (c, 0, 0, -a)^T$.
We now identify the other two eigen-solutions of $M$. Let $v_1 = (0,
1, -1, 0)^T$ and note~that
\[
M v_1 = (e-f) v_1  ,
\]
so $\lambda_1 = (e-f)$ is an eigenvalue. If $e=f$, then the middle
two rows of $M$ are equal and the rank of $M$ is at most 2. (Note
that $\lambda_1$ could be negative, implying that the operator defined
by $M$ is not always positive.)

Now, let $v_2 = (\alpha, 1, 1, \alpha)^T$, where $\alpha$ is a
constant to be determined, and note that
\[
M v_2 =   \left[
\matrix{
\alpha a + 2b + \alpha c \cr
\alpha d + e + f + \alpha\frac{cd}{a} \vspace*{2pt}\cr
\alpha d + e + f + \alpha\frac{cd}{a} \cr
\alpha a + 2b + \alpha c
}
\right ]  .
\]
If $v_2$ is an eigenvector with corresponding eigenvalue $\lambda_2$,
then the first element of $M v_2$ must equal $\alpha\lambda_2$, that
is,
\[
\alpha a + 2b + \alpha c = \alpha\lambda_2  .
\]
Now, using the fact that $2b=1-a-c$, we have
\[
(\alpha-1)(a+c)+1=\alpha\lambda_2  ,
\]
and it follows that
%
\begin{equation}
\label{eqlambda2}
\lambda_2 = \frac{(\alpha-1) (a+c) + 1}{\alpha}  .
\end{equation}
Again, if $v_2$ is an eigenvector with corresponding eigenvalue
$\lambda_2$, then the second element of $M v_2$ must equal
$\lambda_2$, or
\[
\lambda_2 = \alpha d + e + f + \alpha\frac{cd}{a}  .
\]
Now, using the fact that $e=1-d-f-\frac{cd}{a}$, we have
\[
\lambda_2 = \frac{d}{a}(\alpha-1)(a+c)+1  .
\]
Setting our two expressions for $\lambda_2$ equal yields
\[
\alpha d(\alpha-1)(a+c) + a \alpha= a (\alpha-1)(a+c) + a  .
\]
This quadratic in $\alpha$ has two roots: $\alpha=1$ and
\[
\alpha= \frac{a(a+c-1)}{d(a+c)}  .
\]
The second solution is negative and corresponds to a nontrivial
eigenvector. The corresponding eigenvalue~is
\[
\lambda_2 = \frac{1}{a}(a+c)(a-d)  .
\]
If $a=d$, then the sum of the middle two rows of $M$ is equal to twice
the first row.
\end{appendix}

\section*{Acknowledgments}
The third author spoke at length with Professor Richard Tweedie about
the convergence rate of the MDA algorithm during a visit to Colorado
State University in 1993. Although the present work is not directly
related to those conversations, the third author wants to
acknowledge
here his admiration for Professor Tweedie's insights and his gratitude
for his support. The first author's work was supported by NSF Grant
DMS-08-05860. The third author's work was supported by Agence
Nationale de la Recherche (ANR, 212, rue de Bercy 75012 Paris) through
the 2009-2012 project ANR-08-BLAN-0218 Big'MC. The first author
thanks the Universit\'{e} Paris Dauphine for partial travel support
that funded visits to\vadjust{\goodbreak} Paris in 2008 and 2009. The second author
thanks the Agence Nationale de la Recherche through the 2005--2009
project Ecosstat for support that funded a visit to Paris in 2008.
Finally, the authors thank three anonymous reviewers for helpful
comments and suggestions.

%

\end{document}